\documentclass[11pt,tightenlines,twocolumn,superscriptaddress]{revtex4}

\usepackage{amssymb}
\usepackage{color,graphicx,graphpap}
\usepackage{amsmath}
\usepackage{amsbsy}
\usepackage{amsthm}
\usepackage{bbm}
\usepackage{bm}
\usepackage{epsfig}
\usepackage{lscape}
\usepackage{float}
\usepackage{subfigure}

\usepackage{setspace}
\usepackage{hyperref}
\usepackage{upgreek}
\usepackage{xfrac}

\newcommand{\beq}{\begin{eqnarray}}
\newcommand{\eeq}{\end{eqnarray}}

\newcommand{\bmp}{\noindent\begin{minipage}{16cm}}
\newcommand{\emp}{\end{minipage}\vskip 7mm} % 7mm untightened

\def\drawbox#1#2{\hrule height#2pt
        \hbox{\vrule width#2pt height#1pt \kern#1pt
              \vrule width#2pt}
              \hrule height#2pt}

\def\Asym#1#2{\vcenter{\vbox{\drawbox{#1}{#2}
              \kern-#2pt % line up boxes
              \drawbox{#1}{#2}}}}

\def\simge{\mathrel{%
   \rlap{\raise 0.511ex \hbox{$>$}}{\lower 0.511ex \hbox{$\sim$}}}}

\def\simle{\mathrel{
   \rlap{\raise 0.511ex \hbox{$<$}}{\lower 0.511ex \hbox{$\sim$}}}}

\def\s#1{\setbox0=\hbox{$#1$}%
\rlap{\ifdim\wd0>.7em\kern.22\wd0\else\kern.1\wd0\fi /}#1}

\newcommand{\sigmatotal}{\sigma_{\rm eff}} % total effective cross-section for extended particle

\newcommand\ie{\textit{i.e.}\ }

\newcommand{\be}{\begin{equation}}
\newcommand{\ee}{\end{equation}}
\newcommand{\bea}{\begin{eqnarray}}
\newcommand{\eea}{\end{eqnarray}}

\newcommand\scat{\sigma}  %%scattering cross-section
\newcommand\sann{\sigma_{\rm ann}}  %%annihilation cross-section
\newcommand\mc{m_\chi} %%mass of the chi particle
\newcommand\xfo{x_{\rm FO}} %%freeze-out x
\renewcommand\qed{\alpha_{\rm QED}}

\begin{document}

\title{On the Existence of Low-Mass Dark Matter and its Direct Detection}

\author{James Bateman}
\email{jbateman@soton.ac.uk}
\affiliation{Quantum, Light and Matter,} %%to return to just dept address comment out group affiliations & comment in the last affiliation
\author{Ian McHardy}
\email{imh@soton.ac.uk}
\affiliation{Astronomy,}
\author{Alexander~Merle}
\email{A.Merle@soton.ac.uk}
\author{Tim R. Morris}
\email{T.R.Morris@soton.ac.uk}
\affiliation{High Energy Physics Theory,\\ $ $\\ Physics and Astronomy, University of Southampton, Southampton SO17 1BJ, United Kingdom.}
\author{Hendrik Ulbricht}
\email{h.ulbricht@soton.ac.uk}
\affiliation{Quantum, Light and Matter,}
%\affiliation{Physics and Astronomy, University of Southampton, Southampton SO17 %1BJ, United Kingdom {\bf [AM: Should we highlight the different groups?!?]}}

\maketitle

\noindent {\bf Dark Matter (DM) is an elusive form of matter which has been postulated to explain astronomical observations through its gravitational effects on stars and galaxies, gravitational lensing of light around these, and through its imprint on the Cosmic Microwave Background (CMB)~\cite{Ade:2013zuv}. This indirect evidence implies that DM accounts for as much as 84.5$\boldsymbol{\%}$ of all matter in our Universe, yet it has so far evaded all attempts at direct detection~\cite{Kopp:2009qt}, leaving such confirmation and the consequent discovery of its nature as one of the biggest challenges in modern physics. Here we present a novel form of low-mass DM $\boldsymbol{\chi}$ that would have been missed by \emph{all} experiments so far~\cite{Kopp:2009qt,Cirelli:2012tf,ATLAS:2012ky,Chatrchyan:2012me}. While its large interaction strength might at first seem unlikely, neither constraints from particle physics nor cosmological/astronomical observations are sufficient to rule out this type of DM, and it motivates our proposal for direct detection by optomechanics technology which should soon be within reach, namely, through the precise position measurement of a levitated mesoscopic particle~\cite{gieseler2012subkelvin} which will be perturbed by elastic collisions with $\boldsymbol{\chi}$ particles. We show that a recently proposed nanoparticle matter-wave interferometer~\cite{kaltenbaek2012macroscopic}, originally conceived for tests of the quantum superposition principle, is sensitive to these collisions, too.}

Dark Matter interacts at most weakly with ordinary matter. Most theories propose cross-sections for collisions of DM with nucleons which are typically very small, and therefore experimental attempts for its direct detection are usually performed with huge volumes containing many ordinary matter particles. Various different types of particles are discussed as candidates for DM. Very recent attempts to directly observe generic candidates such as supersymmetric DM~\cite{Griest:1988ma} or Kaluza--Klein DM~\cite{Servant:2002aq} did not reach conclusive results and it seems that DM still evades direct observation~\cite{Kopp:2009qt}. Also indirect experiments~\cite{Cirelli:2012tf} searching for annihilation products of DM or attempts to produce DM at the LHC at CERN~\cite{ATLAS:2012ky,Chatrchyan:2012me} have thus far not reported a clear signal, suggesting that WIMPs (\underline{W}eakly \underline{I}nteracting \underline{M}assive \underline{P}article\underline{s})~\cite{Bertone:2010zza}, while being the natural guess for DM, might not exist in nature.  

Alternative and typically very light DM candidates, such as axions~\cite{Weinberg:1977ma,Wilczek:1977pj} or keV sterile neutrinos~\cite{Kusenko:2009up,Merle:2013gea}, have been considered. Often, such particles decay very slowly or annihilate and so produce monoenergetic X-ray photons, which could be regarded as a smoking gun signature for such a type of DM. While dedicated satellite experiments have mostly derived strong limits, a recent detection of a line signal at $3.6$~keV~\cite{Bulbul:2014sua,Boyarsky:2014jta} has attracted the attention of the community but would need to be solidified before a discovery was claimed. Further references on these matters are presented in the supplementary material~\cite{supplement}.

\begin{figure*}[t]
% minipage mit (Blind-)Text
	\begin{minipage}[b]{0.4\textwidth}
	\centering
	\subfigure{\bf a}{\hspace{0.2cm} \includegraphics[width=0.7\textwidth]{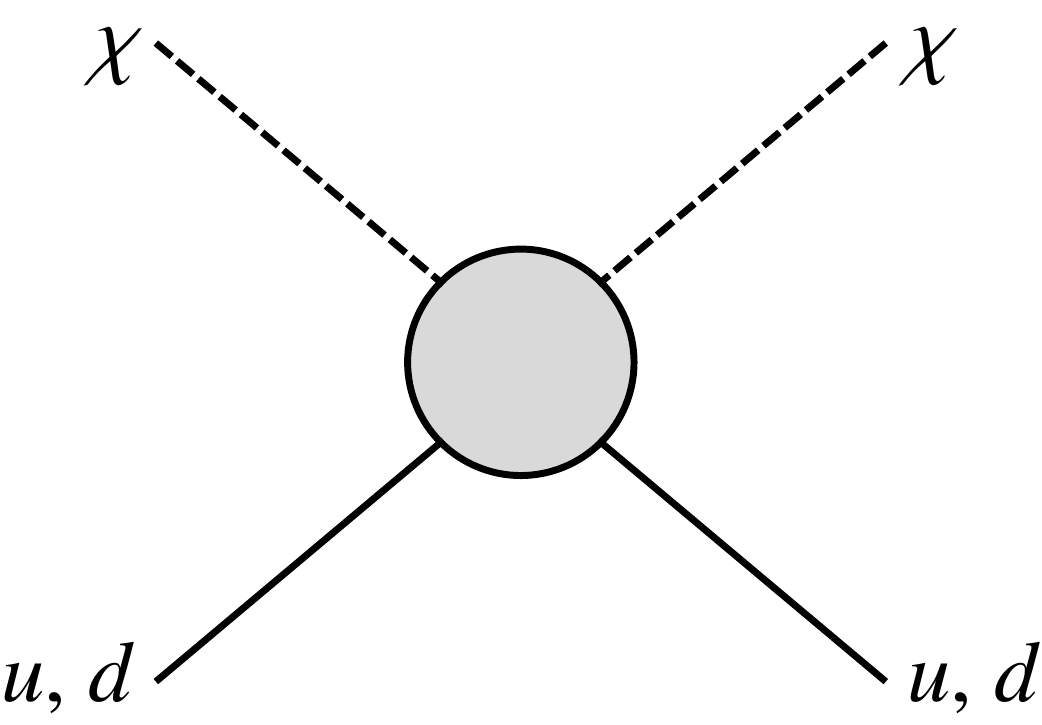}} \\
         \subfigure{\bf b}{\includegraphics[width=0.9\textwidth]{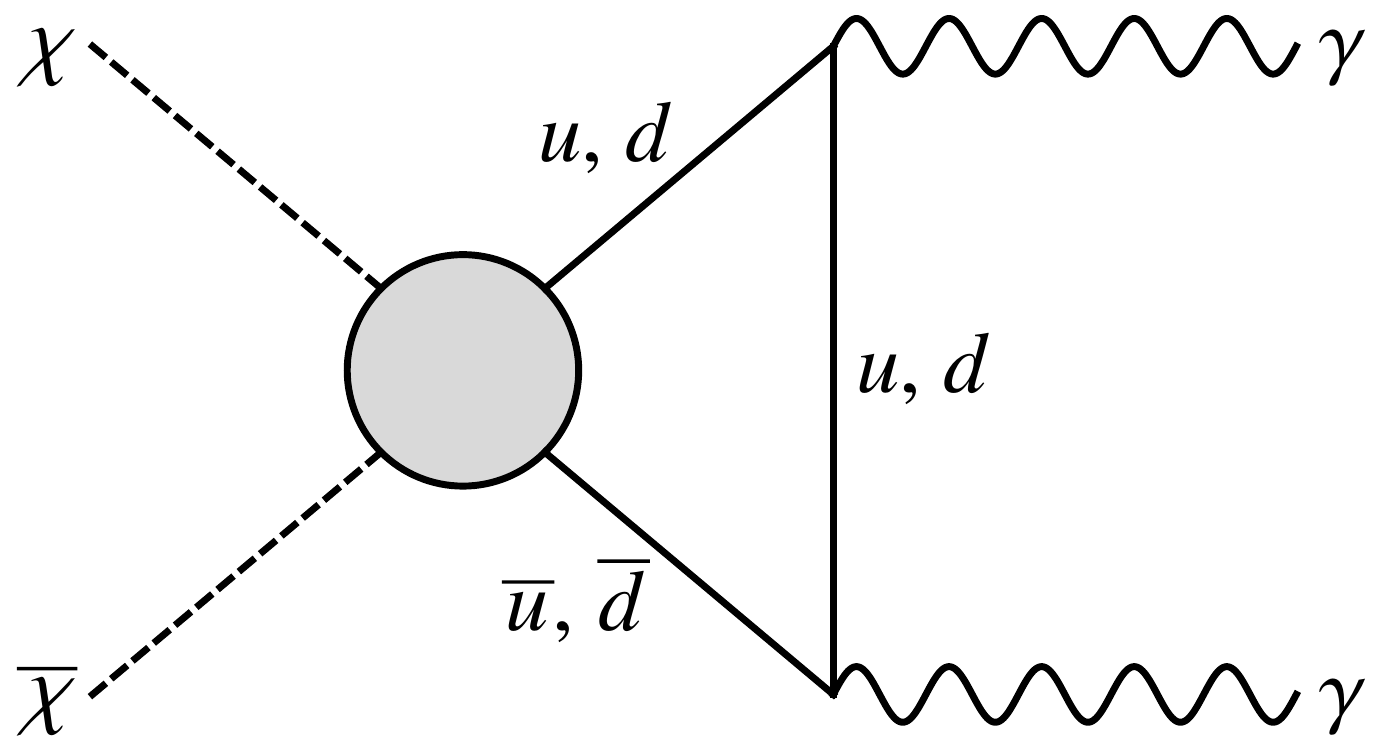}}
	\end{minipage}
	% Auffüllen des Zwischenraums
	\hfill
	% minipage mit Grafik
	\begin{minipage}[b]{0.55\textwidth}
	% \textwidth bezieht sich nun auf die Minipage
	\hspace{-1cm}\subfigure{\bf c}{\includegraphics[width=1\textwidth]{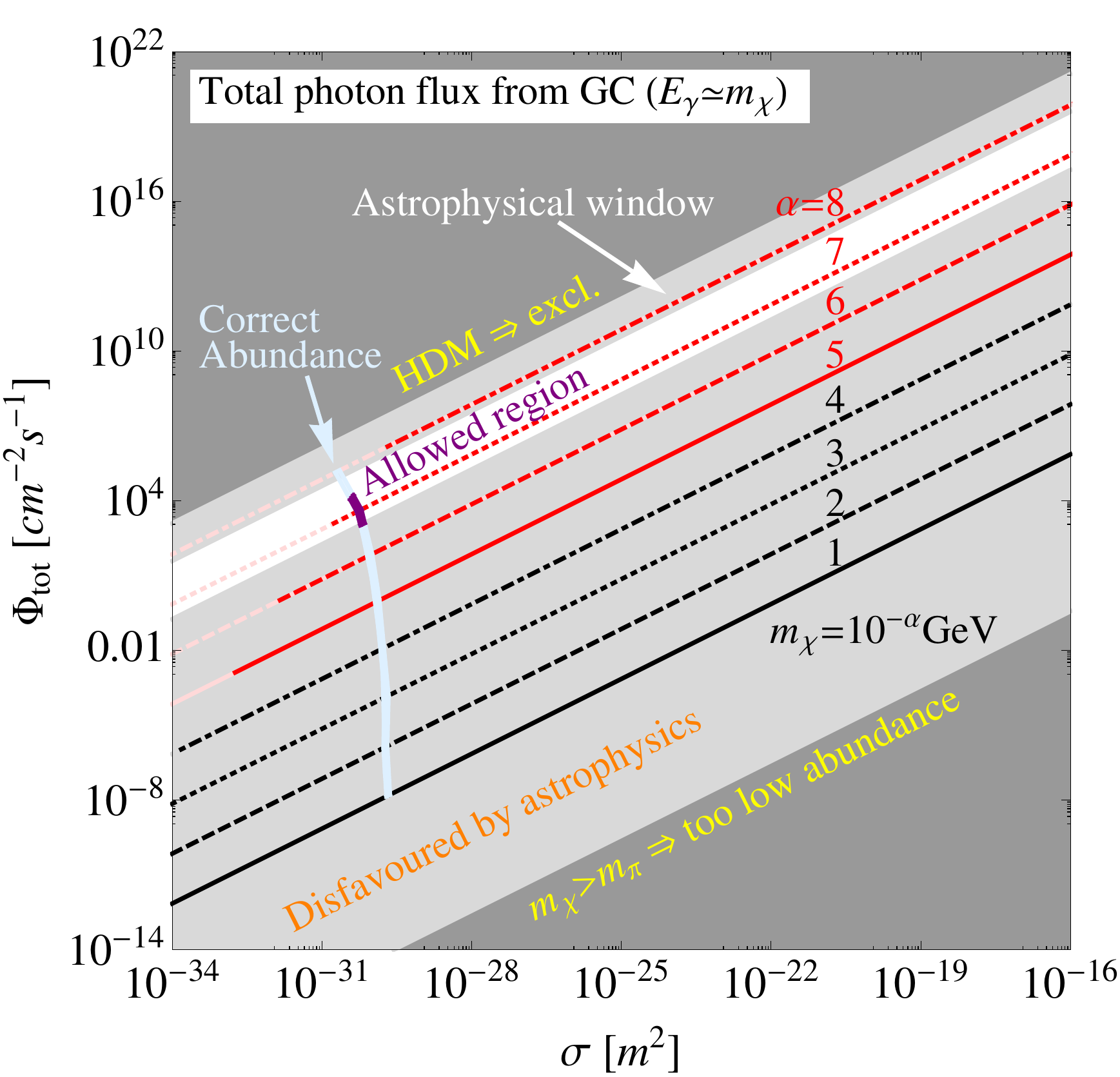}}
	\end{minipage}
\caption{\label{fig:cosmo}\textbf{a}, Feynman diagram relevant for elastic scattering in a test particle. \textbf{b}, Related diagram relevant for DM annihilation in the early Universe. \textbf{c}, Integrated photon flux from the Galactic centre (if not shielded) versus collisional DM cross-section per nucleon, including all constraints which are applicable. The dark gray areas are excluded by consistency arguments, \emph{i.e.}, the DM would either be hot for any choice of parameters (upper left triangle) or its mass would be too large for the suppression mechanism to apply to the annihilation diagram (lower right triangle). The light gray shaded regions are strongly constrained by astrophysical non-observations of the corresponding photons (see supplementary material~\cite{supplement}), although some narrow line signals at particular energies may be difficult to fully exclude. The white patch is allowed by astrophysics. Each point on the red or black lines correspond to a certain DM abundance (see FIG.~6 in the supplementary material~\cite{supplement}), but the parts drawn in light colours would lead to hot DM scenarios, which are excluded as well. The region where the correct amount of DM is produced is marked by the light blue stripe, and the final resulting region allowed by all constraints is drawn in purple. This is what leads us to conclude that, putting all possible constraints together, the mass and scattering cross section of the $\chi$ particle should be around $m_\chi \approx 100$~eV and $\sigma \approx 5\cdot 10^{-31}~{\rm m^2}$.}
\end{figure*}

This work is inspired by a recent suggestion that decoherence in matter-wave interferometry~\cite{arndt2014testing,juffmann2013experimental} could be used as a sensitive detector for very light DM particles~\cite{Riedel:2012ur}. While much of the parameter space is excluded directly or indirectly by existing observations, we find a small range in which such a particle could exist and, with the properties so constrained, we make quantitative predictions for the expected decoherence. Such unorthodox suggestions are crucial to catalyse discussions between disparate areas of physics and facilitate progress in DM searches.

\vspace{2 mm}
\noindent {\bf Cosmological considerations.}  The decisive questions for a concrete DM candidate particle $\chi$ are whether it can be produced in the right amount in the early Universe and whether its velocity spectrum is not too warm to cause problems with cosmological structure formation. In Ref.~\cite{Riedel:2012ur}, the concrete DM candidate was not specified, but putting the constraints from all sides together, we can narrow the possibilities down to a scalar particle $\chi$ with a mass of order $m_\chi \approx 100$~eV and an elastic scattering cross-section on nuclei of $\sigma\approx 5\cdot 10^{-31}~{\rm m}^2$.

The standard process for DM production is \emph{thermal freeze-out}~\cite{supplement}. Cross-sections as needed for a detection in a matter-wave experiment~\cite{Riedel:2012ur} would normally imply far too large annihilation, such that all such DM would be absent today. However, due to its small mass, the $\chi$ particle can only annihilate into \emph{photons} at low temperatures, and this process is intimately connected to, but suppressed with respect to, the direct detection process, \emph{cf.}\ FIGs.~\ref{fig:cosmo}~{\bf a,b}. Thus we can estimate the annihilation cross-section $\sigma_{\rm ann}$ of the DM particle in terms of the detection cross-section $\sigma$ as $\sigma_{\rm ann} v = a + b \langle v^2 \rangle + \mathcal{O}(v^4)$, where $a\sim G^2 m_\chi^2/(4\pi)$, $b \sim a/24$, and $G^2 \sim \alpha_{\rm QED}^2 \sigma v_0 /(18 \pi m_\chi^2)$, with $v$ being measured in units of the speed of light $c$, $v_0 \sim 10^{-3}$, and $\alpha_{\rm QED} \simeq 1/137$. The additional loop-suppression of the annihilation keeps the DM abundance large enough to be consistent with observations.

\vspace{2 mm}
\noindent {\bf Particle Physics and Astrophysical constraints.} The requirements for the $Z^0$-boson and the neutral pion $\pi^0$ not to decay into pairs $\chi \bar{\chi}$ (and the Fermi pressure related Tremaine--Gunn bound~\cite{Tremaine:1979we} for very light fermionic DM) force the particle to have vanishing spin (\emph{i.e.}\ it is a scalar), and the requirement of DM not to be hot excludes very light DM masses, below $10$~eV. The most obvious constraints come from missing energy signatures in colliders~\footnote{Strictly speaking, the energy may not necessarily be missing since our DM particle does interact considerably with nucleons. However, such a signal in a calorimetric experiment could easily be confused with other particles in jets.}, in particular by the smoking gun signature of having a single photon in addition. Several detectors at LHC or previous experiments have reported strong bounds on such a signal~\cite{Aaltonen:2008hh,Abazov:2008kp,CMS:2011tva}. However, using relatively general arguments about the ultraviolet completion behind the effective vertex displayed in FIGs.~\ref{fig:cosmo}~{\bf a,b}, one can see that these high energy bounds do not necessarily have to constrain the low energy vertex needed for DM~\cite{supplement}~\footnote{The production of the $\chi$ particles is suppressed in the $s$-channel, or otherwise hidden in jets. Even though the interaction can in this sense be practically switched off at high energies, this does not affect early Universe cosmology since it does not matter whether the DM particle has been in equilibrium all along or has only entered equilibrium at some point before the freeze-out, as long as it is equilibrated for a sufficiently long time.}.

The particle under consideration nevertheless has a comparatively large collisional cross-section $\sigma$, which  means that it may be absorbed or reflected by the Earth's atmosphere~\cite{supplement}. Furthermore, this could potentially lead to an additional mass contribution for celestial bodies (if the force between DM and ordinary matter is attractive) or to shifts in their trajectories (including precessions). Taking into account that the local DM energy density is tiny, only about $0.4~{\rm GeV}/{\rm cm}^3$, compared to the density of ordinary matter in a typical planet or star, the resulting mass shifts are tiny. For example, the Earth would collect about 1000~tonnes per year (which is a fractional increase of $10^{-19}$ in its mass per year). The $\chi$ DM pressure is of order $P=40$~pPa. In our solar system, the Sun (and Jupiter) would be most affected by the resultant force, but it leads to the negligible acceleration $\sim4\cdot10^{-23}$~ms$^{-2}$. The order of magnitude (in radians) of precessional effects on the planets is given by the ratio of this DM force to the force from the Sun and, for example, for Earth this is an unmeasurably small $\delta\theta\sim3\cdot10^{-17}$ degrees/orbit. Finally, strong constraints arise from a potential annihilation signal arising from the same diagram as the production in the early Universe, \emph{cf.}\ FIG.~\ref{fig:cosmo}~{\bf b}; however, the known observational bounds from several Earth- and space-based telescopes leave a window in which our DM candidate could still live. A detailed discussion can be found in the supplementary material~\cite{supplement}. Thus, all astrophysical constraints are avoided naturally, and all constraints from particle physics do not apply (at least under relatively generic assumptions).

\begin{figure*}[t!]
\subfigure{\bf a} {\includegraphics[width=0.47\textwidth]{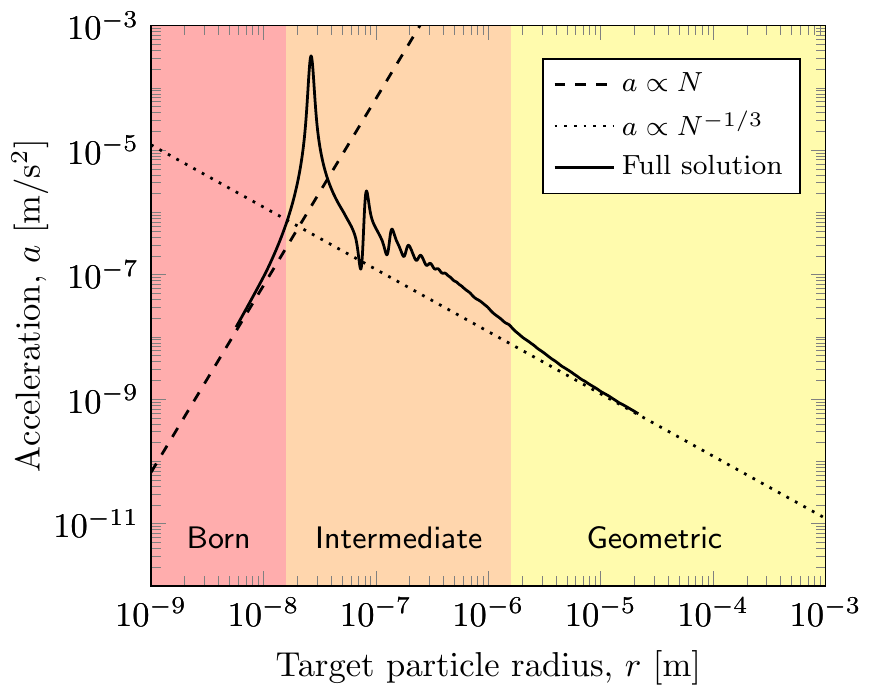}}
\subfigure{\bf b} {\includegraphics[width=0.45\textwidth]{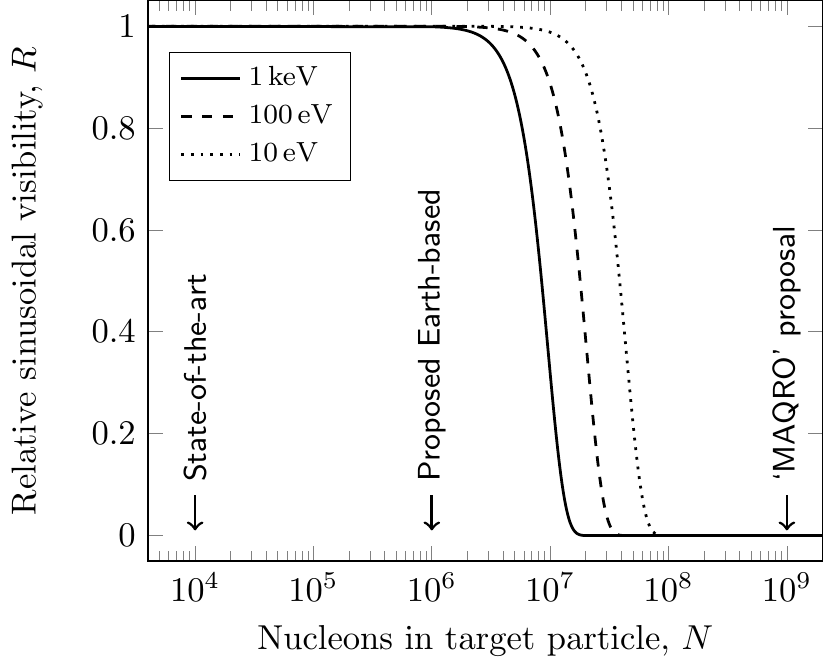}}
\caption{\textbf{a}, Acceleration of a silicon test particle (nucleon number density $1.4\cdot 10^{30}\,\text{m}^{-3}$) across the size regimes for $\chi$ de Broglie wavelength $\lambdabar=1\,\upmu\text{m}$.  For small particles ($r\ll\lambdabar$), the Born approximation holds and acceleration is proportional to nucleon number; for large particles ($r\gg\lambdabar$), the force is proportional to projected area and thus increases slower than the inertia.  In the intermediate regime ($r\sim\lambdabar$), acceleration depends strongly upon the particle shape: for illustration we have chosen a spherical particle with an attractive interaction; the repulsive case is similar. Resonances, which distract from the main argument, have been smoothed by a few times their width. Similar plots are obtained for other de Broglie wavelengths, and the limiting cases are unaffected. \textbf{b}, Reduction in sinusoidal fringe visibility due to elastic collisions for a range of $m_\chi$. Experiments with a similar geometry and path separation are indicated: state-of-the-art experiments have demonstrated $10^4$~\cite{eibenberger2013matter}; an experiment with $10^6$ is proposed~\cite{Bateman:2013near}; and space-based `MAQRO'~\cite{kaltenbaek2012macroscopic} will span the necessary range. For $N\gtrsim 4\cdot 10^7$, the Born approximation for scattering $\chi$ particles is not well satisfied and further theoretical work is needed to fully describe the decoherence.}
\label{fig:accn_and_decoh}
\end{figure*}

\vspace{2 mm}
\noindent{\bf Detection via elastic scattering.} The presence of $\chi$ particles can be detected by the momentum they impart to a test particle via elastic collisions with the constituent nucleons; this recoil is measurable in either a classical detection scheme or via the reduction in fringe visibility in a matter-wave interferometer. Our candidate DM particles are sufficiently light and numerous that we do not expect to resolve individual scattering events; rather, we expect an overall drift in the direction of the DM, and a very small Brownian-like diffusion.

{\it Scaling with target particle size:} 
A consequence of the low $\chi$ mass is that the de Broglie wavelength $\lambdabar=\lambda/(2\pi)$ is large compared to the internuclear separation in normal matter:  $\lambdabar\gtrsim 100\,\text{nm}$.  The $\chi$ hence scatters \emph{coherently} from the constituent nuclei. For small particles under the Born approximation, all nuclei are subject to the same field from the incident $\chi$, and we find an effective cross-section $\sigma_\text{eff}=\sigma N^2$. Conversely, for large particles, the flux is attenuated and the cross-section is the projected surface area $\sigma_\text{eff}\propto N^{2/3}$. In the intermediate regime, details of the interaction depend strongly on particle shape and on whether the underlying interaction is attractive or repulsive. For illustration, we consider a spherical particle with an attractive potential and we calculate the interaction via partial waves~\cite{supplement}; the expected acceleration $a=\sigma_\text{eff}P/M$, where $M$ is the particle mass, as shown in FIG.~\ref{fig:accn_and_decoh} {\bf a}, reduces to the Born approximation and to the geometrical approximation in the respective limits. Details of size-dependent acceleration in the intermediate regime, if observed, will allow for an independent measurement of the $\chi$ DM pressure $P$ and collisional cross-section $\sigma$.

{\it Dark Matter optics:} For macroscopic objects, $\chi$ particles experience an average potential and, in close analogy with neutron optics~\cite{sears1989neutron}, the interaction may be described using a refractive index $\eta=\sqrt{1-(\lambda/\lambda_\text{c})^2}$, where we identify the `critical wavelength' $\lambda_\text{c}=\sqrt{{\pi}/{n a_s}}$, with $n$ being the number-density of nucleons in the material, and the scattering length $a_s=\pm0.2\,\text{fm}$ is found via the low-energy limit in which $\sigma=4\pi a_s^2$.  The uncertainty in sign (and thus whether $\lambda_\text{c}$ is real or imaginary) arises because  the cross-section is insensitive to whether the underlying interaction is attractive ($-$) or repulsive ($+$). For typical materials, $\left|\lambda_\text{c}\right|\approx 100~\text{nm}\ll\lambda$, and we expect $\chi$ particles to be strongly reflected.

\begin{figure}[t]
\includegraphics[width=0.45\textwidth]{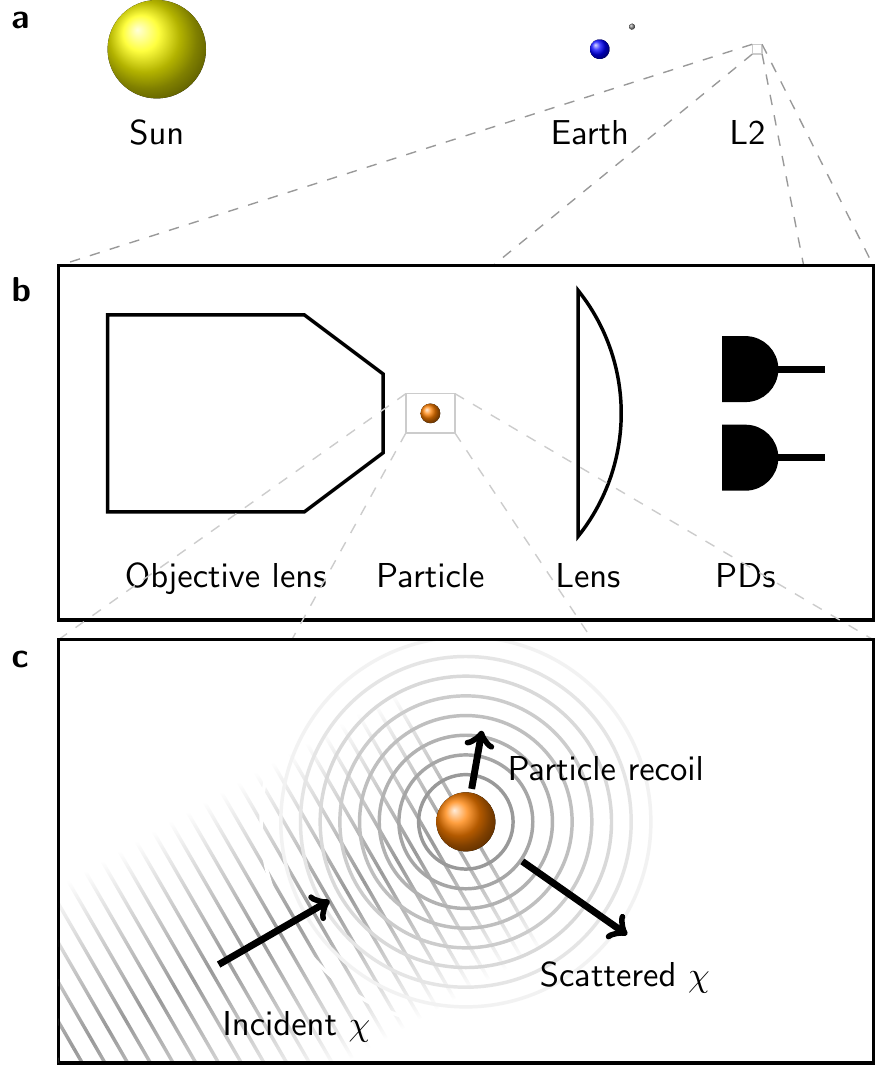}
\caption{Illustration of the suggested experiment, the hardware for which can be provided by the proposed `MAQRO' space-craft~\cite{kaltenbaek2012macroscopic}. \textbf{a}, Location of the space-craft at Lagrange point 2 in the context of our solar system (not to scale). \textbf{b}, Close-up of the optical arrangement: a compound objective lens provides high numerical aperture focusing for laser light to create a gradient-force dipole trap for a micron-scale particle. Light, which diverges strongly after the particle, is collected by a lens. Interference between the laser light and the light scattered coherently by the particle gives rise to a difference in intensity across the cross-section which, when measured by balanced photodiodes (PDs), provides sub-wavelength position information in three dimensions~\cite{gieseler2012subkelvin}. \textbf{c}, A further close-up, showing $s$-wave scattering of a $\chi$ DM particle, with an approximately plane-wave incident wavefunction and an example scattering outgoing direction with the associated recoil of the test particle~\cite{supplement}.}
\label{fig:schematic}
\end{figure}

\vspace{2 mm}
{\it Acceleration of a mesoscopic particle: } Given the possibility of a measurable effect upon nanometre-sized particles, and the uncertainty about whether $\chi$ particles will penetrate the Earth's atmosphere, we propose a space-based experiment, as illustrated in FIG.~\ref{fig:schematic}. Particle radii in the range $10~\text{nm}\leq r\leq 1~\upmu\text{m}$ are expected to show accelerations $a\gtrsim 0.1~\upmu\text{m}/\text{s}^2$, with possibly much higher values and a rich size-dependent structure. Recently, 140 nm particles have been held in vacuum in a $120\,\text{kHz}$ harmonic trap provided by a tight laser focus and feedback `cooled' to reduce the uncertainty in both their position ($<$1 nm) and velocity (500 $\upmu\text{m}/$s)~\cite{gieseler2012subkelvin}. For a thermal state, the velocity uncertainty is the product of trap frequency and position uncertainty and, in ultra-high vacuum where gas collisions are negligible, one may decrease the trap frequency considerably; for a $10\,\text{kHz}$ trap frequency, we expect a velocity uncertainty below $50\,\upmu\text{m}/\text{s}$. After several minutes of free-flight under these conditions, the positional uncertainty will be sub-millimetre while acceleration from collisions with $\chi$ particles will give a millimetre-sized displacement. The effect is also observable without any such improvements; the displacement will be revealed in the statistics of position measurements.

{\it Matter-wave decoherence: }
The prediction of an acceleration is based upon the assumption that the Earth moves through the local DM distribution at some appreciable speed.  However this local distribution is uncertain particularly for this yet-to-be-simulated DM candidate, so here we propose a detection scheme which does not rely on some overall drift. 

Elastic scattering events can be interpreted as revealing partial which-way information or, in a more complete treatment including recoil, diffusing momentum in a quantum Brownian Motion~\cite{joos2003decoherence}. While individual collisions may not affect the visibility significantly, many such events will have a measurable effect. A proposed space-based matter-wave nanoparticle interferometer~\cite{kaltenbaek2012macroscopic} will be sensitive to this decoherence mechanism and the effect can be controllably extinguished by shielding the nanoparticle from the DM flux. We analyse the decoherence for a similar interferometer~\cite{Bateman:2013near}, where a nanoparticle, prepared in a thermal state of a harmonic oscillator via feedback cooling, provides a point-like source for a near-field (Fresnel region) Talbot interferometer using a phase grating of period $\Lambda$ provided by a standing light-wave.

The overall fringe pattern is found via a Wigner function phase-space treatment and is expressed as a Fourier series, the first order of which may be robustly extracted from experimental data by fitting to a sinusoid. Each decoherence mechanism reduces this amplitude by a factor \mbox{$R=\exp\left[-W\,f\left(\tfrac{8}{9}\Lambda\right)\right]$}, where \mbox{$W=cv_0\sigma_{\rm eff}\,\tau\,\rho_\chi/m_\chi$} is the total number of events (the flux multiplied by the cross-section and the duration $\tau$ of the experiment), the numerical factor $\sfrac{8}{9}$ comes from the geometry of the experiment, and $f(x)$ describes the spatial resolution of each event~\cite{supplement}. For $x\ll\lambdabar$, individual events affect the state little and multiple events are necessary to cause a measurable decoherence; for $x\approx\lambdabar$, each event reduces visibility by approximately one half. A silicon particle of internal temperature below $50~\text{K}$ in ultra-high vacuum has negligible decoherence from the two important mechanisms, black-body radiation and gas collisions, and the reduction in visibility is dominated by decoherence from collisions with $\chi$ particles; this is shown in FIG.~\ref{fig:accn_and_decoh}~{\bf b} and we see that this decoherence is significant for experimentally accessible masses.

\vspace{2mm}
\noindent{\bf Concluding remarks.}
We predict a light form of DM and we have argued that it is possible that this specific form of DM, the particle $\chi$, would not have been observed in any experiment so far. We identify the mass range of $\chi$ and its collisional cross-section. We hope that this will catalyse further developments of more detailed particle theories and allow for more precise predictions of the properties. While both of the optomechanical experiments which we have proposed are space-based, the possibility of Earth-based detection is the topic of ongoing research; the prospects for such detection depend on the details of the particle theory, which is yet to be developed, and on the details of $\chi$ particle interaction with the atmosphere. Both the modulation of the DM flux as expected on Earth due to planetary motion and the possibility of extinguishing the flux with a mechanical shutter provide clear experimental signatures for identification of $\chi$. Observation of a size-dependent acceleration would reveal far more details about the nature of these particles.

%Collisions with DM particles as modelled in this work provide a decoherence mechanism which, because scattering results from a coherent interaction, depends strongly upon the size and shape of the target. This suggests the tantalising possibility that this process could explain the phenomenological quantum to classical transition because this DM-mediated decoherence might remain when all other known forms of decoherence have been eliminated through great experimental effort.  However, such decoherence is not well described by current theories and we hope to resolve this question through our ongoing research.

Experimentally, the possibility of defining a refractive index for the interaction of $\chi$ with ordinary matter allows for the implementation of optical elements to manipulate, guide, and even suppress reflections of DM beams. We can hope to greatly increase the local density of this type of DM, if it exists. Furthermore, complementary detection techniques should be studied as well as the possibility for direct $\chi$ production at low energy, high intensity photon colliders~\cite{Marklund:2006my}.

\vspace{2mm}
\noindent{\bf Acknowledgments.}
We thank Markus Arndt, Angelo Bassi, Sasha Belyaev, Tony Bird, Sandro Donadi, Phil Charles, Giulio Gasbarri, Philip Haslinger, and Jess Riedel for discussions.  AM acknowledges support by a Marie Curie Intra-European Fellowship within the 7th European Community Framework Programme FP7-PEOPLE-2011-IEF, contract PIEF-GA-2011-297557, as well as partial support from the European Union FP7 ITN-INVISIBLES (Marie Curie Actions, PITN-GA-2011-289442). HU and JB wish to thank the UK funding agency EPSRC for support under grant (EP/J014664/1), the Foundational Questions Institute (FQXi), and the John F Templeton foundation under grant (39530). TRM and IMcH acknowledge STFC support through Consolidated Grants ST/J000396/1 and ST/J001600/1, respectively.

\onecolumngrid
\section{Supplementary Materials}
Here we give more detailed information and derivations for \emph{On the Existence of Low-Mass Dark Matter and its Direct Detection}.

\subsection{Cosmology of Dark Matter}

As we had explained in the main text, Dark Matter (DM) interacts at most weakly with ordinary matter, which is why experimental attempts for its direct detection are usually performed with huge volumes containing many ordinary matter particles. A generic type of DM would be \underline{W}eakly \underline{I}nteracting \underline{M}assive \underline{P}article\underline{s} (WIMPs), which have masses of a few $100$~GeV and which typically interact with roughly weak interaction strength. Among the generic WIMP candidates are supersymmetric DM particles (\emph{e.g.}\ neutralinos~\cite{Griest:1988ma,Gelmini:1990je,Jungman:1995df}, sneutrinos~\cite{Hall:1997ah,Falk:1994es,Cerdeno:2008ep,Arina:2007tm}, or gravitinos~\cite{Takayama:2000uz,Ellis:2003dn,Steffen:2006hw,Buchmuller:2007ui}) or candidates motivated by extra spatial dimensions (Kaluza-Klein (KK) DM: \emph{e.g.}\ KK-gauge bosons~\cite{Servant:2002aq,Cheng:2002ej,Burnell:2005hm,Bonnevier:2011km,Melbeus:2011gs} or KK-Higgses~\cite{Arrenberg:2008wy,Melbeus:2012wi}). Up to now, direct detection attempts~\cite{Aprile:2012nq,Akerib:2013tjd,Agnese:2013jaa,Archambault:2009sm,Agnese:2014aze}, indirect experiments searching for annihilation products of DM~\cite{Aartsen:2013dxa,Adrian-Martinez:2013ayv,IceCube:2011aj}, or attempts to produce DM at the LHC~\cite{ATLAS:2012ky,Aad:2012fw,Chatrchyan:2012me,Chatrchyan:2012tea} have thus far not reported a clear signal, suggesting that WIMPs, while being the natural guess for DM, might in reality not exist in nature.  

Very light and hardly interacting DM candidates exist, too, such as axions~\cite{Holman:1982tb,Steffen:2008qp,Preskill:1982cy} or keV sterile neutrinos~\cite{Dodelson:1993je,Shi:1998km,Canetti:2012kh,Bezrukov:2009th,Nemevsek:2012cd,King:2012wg,Shaposhnikov:2006xi,Bezrukov:2009yw,Kusenko:2006rh,Petraki:2007gq,Merle:2013wta,Merle:2013gea,Kusenko:2009up}. Such particles might annihilate or (very slowly) decay and could thus lead to monoenergetic X-ray photon signatures. Dedicated satellite experiments have derived strong limits~\cite{Boyarsky:2005us,Boyarsky:2006ag,Watson:2006qb,Boyarsky:2007ay,Boyarsky:2007ge,Loewenstein:2008yi,Boyarsky:2006fg,Watson:2011dw,Loewenstein:2012px}, but recently a detection of a tentative signal at $3.6$~keV has been reported~\cite{Bulbul:2014sua,Boyarsky:2014jta}.

Whichever DM candidate is considered, it has to be demonstrated that it can be produced in the correct amounts on the early Universe and that it escapes all known constraints. In this part of the supplementary material, we recall the standard mechanism behind DM production and put our particle $\chi$ into  context amongst the already hypothesised DM candidates.\\

\paragraph{Particle production in the early Universe: an illustrative sketch}

Early Universe cosmology is a subject by itself, and there exist many excellent textbooks on the subject (Ref.~\cite{Kolb:1990vq} being one example). It is clear to us that this paper could potentially be read by scientists from very different communities, which is why we would like to explain the required basics in some detail -- even though certain readers might already know this. While we cannot review all the theory behind particle production in the early Universe, we at least want to give a snapshot of how things work.

The first point is that at high temperatures, as present in the early Universe, all particles can be regarded as practically massless, \emph{i.e.}\ they effectively act as \emph{radiation}. The other components of the Universe, non-relativistic matter and Dark Energy, are completely negligible at this early stage. All particles $\chi$ with a sufficiently large annihilation cross-section, \emph{i.e.}\ a sufficiently high rate of producing, or being produced by, two photons, $\chi \overline{\chi} \leftrightarrow \gamma \gamma$, are in \emph{thermal} (and also \emph{chemical}) \emph{equilibrium}~\cite{Gondolo:1990dk}. This essentially means that the (thermally averaged) annihilation rate of a particle $\chi$ and its antiparticle $\overline{\chi}$ into photons is exactly the same as the inverse rate, within a certain volume. Hence, what matters is the \emph{density} of the particle species $\chi$. The most important requirement to keep a particle in thermal equilibrium is that its interaction rate with photons (or, more precisely, with SM particles which in turn interact with photons) is large enough. Furthermore, we should note that a massless particle is very easy to produce, since essentially all the photons have a large enough energy to produce the particle $\chi$ as long as $T \gg m_\chi$, \emph{i.e.}\ the temperature of the thermal plasma (and hence the photons) is larger than the mass $m_\chi$ of the particle $\chi$. Thus, the number density of a species $\chi$ in thermal equilibrium will be very large if $\chi$ is highly relativistic, \emph{i.e.}\ $T \gg m_\chi$.

As time goes by, the Universe expands and by this it cools down, due to the associated redshift of all the radiation in the Universe. However, this does not only decrease the temperature but it also slows down the effective interaction rates. These are essentially given by the number density times the thermally averaged cross-section, $n_\chi \langle \sann v \rangle$, and the number density $n_\chi$ decreases while the Universe is expanding. Thus the interaction rate further and further decreases until, at some point, it drops \emph{below} the expansion rate (the Hubble function) $H$ of the Universe. At that point, no particles of the species $\chi$ can be annihilated anymore, because their number density is too small and the particles do not find any interaction partners to annihilate with. This is what is called \emph{thermal freeze-out}. Obviously, the time (or temperature) at which this freeze-out happens depends on the value of the cross-section and can by this be very different, depending on the properties of the species $\chi$. For example, it depends on whether $\chi$ is electrically charged or, more generally, which interactions it participates in. Notably, even if the species $\chi$ is only weakly interacting this is nevertheless by far enough to keep it in thermal equilibrium in the early Universe. The simple reason for this is that the weakness of the weak interactions purely comes from the relatively large mass of the exchanged $W^\pm$ and $Z^0$ bosons, which are also practically massless at high enough temperatures, and the coupling strength itself is comparable to that of electromagnetism. Having explained how a species $\chi$ can undergo thermal freeze-out, we also note that a frozen-out species $\chi$ will survive in the Universe until today if it is stable (or, at least, if it has a lifetime that is considerably larger than the lifetime of the Universe). This is exactly why, often, DM is thought of as being a massive particle that is only weakly interacting: it is not allowed to be electrically charged, since then it would directly couple to photons and hence not be ``dark'', but if it is charged under weak interactions only, it can nevertheless enter thermal equilibrium in the early Universe and then undergo thermal freeze-out.

There is still one more subtlety involved which we have to discuss: freeze-out can happen when the species $\chi$ is highly relativistic or when it is non-relativistic or anywhere in between. We have already mentioned that the number density $n_\chi$ of the particle $\chi$ is comparatively large when the particle is relativistic, because it is easy to produce such particles. However, in the case of $T \ll m_\chi$, only the photons with the highest momenta have enough energy to produce an $\chi \overline{\chi}$ pair (this is the reason why we need the \emph{thermally averaged} cross-section, to take into account the thermal momentum distribution of the particles in the Universe). Then, even though the photon density might still be large, it is a rare event that $\chi \overline{\chi}$-pairs are produced and its number density decreases considerably. In fact, it even decreases exponentially with the inverse temperature, $n_\chi \propto e^{- m_\chi / T}$. This will also influence the thermal freeze-out since, as already mentioned, it is actually the product of the number density and the thermally averaged cross-section, $n_\chi \langle \sann v \rangle$, which has to be compared to the expansion rate $H$ of the Universe. Now we are ready to understand one basic property of thermal freeze-out: \emph{the more relativistic a particle species $\chi$ is at freeze-out, the larger its number density will be}.

Finally, we have to understand that what is ``measured'' or, rather, inferred from the observation of the cosmic microwave background (CMB) is in fact an \emph{energy density}. If we consider a \emph{co-moving volume} (\emph{i.e.}\ a piece of space in a coordinate system that grows together with the expanding Universe), the number density $n_\chi$ of this volume will remain constant after freeze-out, since no $\chi$ particles annihilate or are produced anymore.~\footnote{Strictly speaking, this is not true due to the change in the effective number $g_*$ of relativistic degrees of freedom, so that the quantity which remains precisely constant is not the number density $n_\chi$ but rather the so-called \emph{yield} $Y_\chi = n_\chi / s$, where $s$ is the entropy density at a given temperature. However, this is a slightly abstract quantity and using it does not fundamentally change the general argumentation, which is why we will therefore stick to the use of the number density for the purpose of giving an illustrative explanation.} The $\chi$ particles within this co-moving volume will eventually slow down, such that their velocities at late times become negligible in comparison to the speed of light $c$, and hence the final energy density $\rho_\chi$ that remains in $\chi$ is simply given by the product of their mass and their number density, $\rho_\chi = m_\chi n_\chi$. The quantity that is derived from observations is called the \emph{abundance} $\Omega_{\rm DM}$ (or, more commonly, $\Omega_{\rm DM} h^2$, with $h$ being the reduced Hubble constant), and it is essentially the fraction of the total energy density (actually the so-called \emph{critical density}) of the Universe which resides in DM particles today. The current most up-to-date value comes from the 2013 data release of the Planck satellite~\cite{Ade:2013zuv}, and the observed $1\sigma$ range derived from Planck data only is given by
\be
\label{odm}
\Omega_{\rm DM} h^2 = 0.1199 \pm 0.0027\,.
\ee

Whichever DM candidate we consider, if it is to be the only constituent of the DM observed in the Universe, it is absolutely indispensable for it to reproduce the correct abundance. In any case our species $\chi$ can clearly not exceed the above measured abundance. On top of that, further constraints arise \emph{e.g.}\ from cosmological structure formation, from direct searches, or from indirect bounds, as well as from consistency arguments (the new particle should ``fit'' with the known ones, loosely speaking). We will investigate in the following whether all these conditions can be fulfilled for the type of DM under consideration.\\

\paragraph{Calculating the Dark Matter abundance}

Turning now to DM as needed for Ref.~\cite{Riedel:2012ur}, we first of all want to get some understanding of the properties of the DM candidate under consideration. Probably the first question a particle physicist would ask is about the spin -- is the DM particle a scalar (spin~$0$), is it a fermion (spin~$1/2$ or $3/2$), or is it a vector (spin~$1$), as these are the only possibilities which exist in renormalisable theories. However, in fact, the more important question is about the possible suppression of the annihilation cross-section $\sann v$: is the leading order an $s$-wave (such as typical for scalar, Dirac fermion, and vector DM), which is without velocity dependence, or is a $p$-wave (as typical for Majorana fermions or for cases where certain suppressions apply), which is suppressed by the square of the velocity, $v^2$ (this being small for non-relativistic DM). Unfortunately, we cannot easily answer this question, since the exact properties of the DM candidate are not specified in Ref.~\cite{Riedel:2012ur}. 

On the other hand we want to try to understand what the prospects are for an experiment as suggested in Ref.~\cite{Riedel:2012ur} in general. So, the only way we can proceed is to try a certain generic candidate. For simplicity, we have decided to assume a \emph{scalar particle}, since then the equations look simplest and since the relation between the DM annihilation cross-section and the scattering on the molecules is also easiest in that case. As we will illustrate later, this choice is in fact probably the best one could make. The reason is that, at several places, the scalar DM case will have a tendency to save itself from strong bounds.

The essential consequence of freeze-out is as follows: the \emph{larger} the annihilation cross-section (\emph{e.g.}, if it is unsuppressed) the \emph{smaller} the final abundance will be, due to more of the DM particles annihilating before the freeze-out. Furthermore, the more non-relativistic the particles are at freeze-out the more suppressed their number density and hence their final abundance will be. However, as we had specified earlier, the decisive cross-section is the \emph{annihilation cross-section into SM particles} $\sigma_{\rm ann}$, while what is given in Ref.~\cite{Riedel:2012ur} is the \emph{scattering cross-section on nucleons} $\sigma$, and these two seem to be very different quantities at first sight.

We can try to find an easy estimate. It is well-known that the scattering cross-section $\sigma_{\rm quark}$ on a quark can be estimated from the scattering cross-section $\scat$ on a nucleon as
\begin{equation}
 \sigma_{\rm quark} \approx \frac{\scat}{3^2},
 \label{eq:sigma-estimate_1}
\end{equation}
since every nucleon contains three quarks. (This is in fact the coherence factor championed in Ref.~\cite{Riedel:2012ur} but operating at the sub-nucleon level.) Furthermore, to obtain the annihilation cross-section from the direct detection cross-section, at least for a scalar particle, we do not have to worry about spin-dependent scattering contributions and can optimistically try to rotate the Feynman diagram by $90^\circ$ to estimate
\begin{equation}
 \sigma_{\rm ann,\ naive} \gtrsim 2 \sigma_{\rm quark} = \frac{2}{9} \scat,
 \label{eq:sigma-estimate_2}
\end{equation}
where the factor 2 takes into account the fact that the DM particle can interact with both, $u$- and $d$-quarks. This is an important observation: the $u$- and $d$-quarks contained in the nucleons inside the experiment are \emph{precisely the same particles} as present in the early Universe, which leads to the annihilation diagram displayed in FIG.~1~{\bf b} (main text).

Of course, there is no principal reason that the DM particle $\chi$ could not interact with even more types of quarks or even further SM particles, hence the ``$>$'' in Eq.~\eqref{eq:sigma-estimate_2}. That would \emph{increase} the annihilation cross-section and therefore make the resulting abundance smaller. Indeed, for many popular DM candidates the annihilation cross-section is indeed considerably larger than the direct detection cross-section~\cite{Kakizaki:2005uy,Cheng:2002ej,Melbeus:2012wi,Honorez:2010re,Hooper:2005fj,Belanger:2007dx,Servant:2002hb,Servant:2002aq}, even though examples for the contrary case exist as well~\cite{Chang:2013oia}, and suppressions as \emph{e.g.}\ for Majorana DM can apply. For simplicity, we stick to the minimal assumption that the DM only interacts with the quarks present in the proposed quantum decoherence experiment. Indeed, the estimate in Eq.~\eqref{eq:sigma-estimate_2} comprises the minimal assumption on the cross-section. 

However, there is one subtlety to discuss. As soon as the temperature of the Universe (or, rather, the temperature of the thermal plasma) falls below the QCD confinement scale between 150 and 450~MeV, $u$- and $d$-quarks do not anymore exist as free particles. Instead, the lightest QCD bound states will exist, namely pions with masses of around $150$~MeV~\cite{Beringer:1900zz}. However, in the mass range under consideration (where $m_\chi < 0.1$~GeV), our DM particles are not able to annihilate into pions at rest for kinematical reasons, since $2 m_\chi < 2 m_\pi$. Thus, the only annihlation channel which will be open at that stage will be the annihilation into two photons, $\chi \overline{\chi} \to \gamma \gamma$. The corresponding Feynman diagram contains a loop, \emph{cf.}\ FIG.~1~{\bf b} in the main text, which means that it is suppressed by a factor of roughly $\qed^2/(16 \pi^2) \sim 3\cdot 10^{-7}$, where $\qed \simeq 1/137$ is the fine structure constant, compared to the annihilation cross-section into quarks. Thus, we must correct Eq.~\eqref{eq:sigma-estimate_2} by this suppression factor to obtain a reliable estimate,
\begin{equation}
 (\sann v)_{\rm refined} \approx \frac{2}{9} v \frac{\qed^2}{16 \pi^2} \scat,
 \label{eq:sigma-estimate_3}
\end{equation}
where we have already included the thermal velocity $v = \sqrt{\frac{3 T}{m}}$ for a particle with mass $m$ and temperature $T$, as generically obtained from the thermal distributions in the early Universe.~\footnote{Note that, as usual in particle physics and cosmology, $v$ is dimensionless, \ie we are here using units in which $c=1$.} This cross-section is not decisive for high temperatures, where quarks can be produced without problems, but late enough in the evolution of the Universe -- which is exactly the time that is decisive for our type of DM -- it is the only one which is there. This leads to a sudden drop in the cross-section, however, since the cross-sections under consideration are still comparatively high, even this suppressed annihilation rate is enough to keep the DM particles in thermal equilibrium for some time. But at some point the DM particles will nevertheless freeze out due to the suppressed cross-section, which is the reason for their sizable final abundance. This is the first instance where the DM under consideration saves itself from a disaster: typically, a DM particle with such large cross-sections on quarks would stay in equilibrium for a too long time, eventually become very non-relativistic, and thus have a strongly suppressed final abundance. However, due to the additional kinematical constraint and consequent loop suppression of the cross-section, this does not happen and we are left with a large enough abundance.

Note that we could have obtained the above result also in another way: we could assume an effective interaction Lagrangian between the DM and protons.~\footnote{In principle, we could also assume a similar interaction with neutrons, which would change the resulting equation by a factor of 2 as we will explain.} The simplest such interaction (scalar type, with strength $\xi$) would be given by the Lagrangian
\begin{equation}
 \mathcal{L}_{\rm eff} = \xi |\chi|^2 \overline{p} p,
 \label{eq:eff_nucleon_int}
\end{equation}
which results into a Feynman rule $(-i)\xi$. Using textbook methods, it is easy to show that the scattering cross-section of non-relativistic DM on nucleons would then be given by $\scat \simeq \frac{\xi^2}{4\pi}$, while the (hypothetical, due to the too small mass) annihilation into a proton-antiproton pair would be given by $(\sann v)_{\chi \chi \to \overline{p} p} \simeq \frac{\xi^2}{4\pi} \sqrt{1-\frac{m_\chi^2}{E_\chi^2}}$, for an initial energy $E_\chi$ of $\chi$. The square root is nothing else than the velocity of the initial state $\chi$ and the prefactor is nothing else than $\scat$. However, we again have to take into account that non-relativistic annihilation of $\chi$ is only possible into photons, which leads to the analogous graphs as in FIGs.~1~{\bf a,b} (main text) now with protons, and thus
 the same suppression factor $\qed^2/(16 \pi^2)$ as above. Taking into account that only protons can couple to photons, while neutrons cannot, could impose an additional suppression of $1/2$, but under the assumption in Eq.~\eqref{eq:eff_nucleon_int} that $\chi$ only couples to protons, the resulting estimate is
\begin{equation}
 (\sann v)_{\chi \chi \to \overline{p} p} \simeq \frac{\qed^2}{16 \pi^2} \scat v,
 \label{eq:ann_estimate}
\end{equation}
which is in good agreement with the above estimate using quarks as internal states. This reasoning in particular justifies the argument of ``rotating'' the Feynman diagram by $90^\circ$.

The next point to discuss is the variation of the annihilation cross-section with the temperature. As explained before, the decisive quantity is in fact the thermally averaged cross-section times the velocity, $\langle \sann v \rangle$, which we will approximate by Eq.~\eqref{eq:sigma-estimate_3} for low temperatures (this will allow us to estimate the interaction strength). Depending on the temperature $T$ and the mass $m_\chi$, the thermal average can be more or less decisive. A particularly delicate region is the one where the particle is neither highly relativistic (``hot'') nor fully non-relativistic (``cold'') at the time of the freeze-out, but somewhere in between (``warm''). This region looks very different depending on whether the annihilation cross-section is dominated by $s$- or $p$-wave contributions. We again have some freedom here and we have decided to assume an $s$-wave contribution, as generic for a scalar DM particle. Then we can make use of Ref.~\cite{Drees:2009bi}, where an easy and relatively accurate interpolation formula between the hot, warm, and cold regions had been suggested:
\begin{equation}
 \langle \sann v \rangle_{\rm approx} \approx \frac{G^2 m_\chi^2}{16 \pi} \left( \frac{12}{x^2} + \frac{5+4x}{1+x} \right),
 \label{eq:inter}
\end{equation}
where $G$ is an effective coupling constant~\footnote{Note that, in Ref.~\cite{Drees:2009bi}, $G$ is in fact an effective 4-\emph{fermion} coupling, which we sloppily use to describe an interaction between two \emph{scalar} DM particles with two quarks, which are fermions. This may look like a major flaw at first sight. However, since we only use the coupling $G$ as a translation between Eq.~\eqref{eq:sigma-estimate_2} and the non-relativistic limit of Eq.~\eqref{eq:inter}, its mass dimensions do in fact not play any role.} and $x \equiv m_\chi / T$ is the usual variable which essentially describes the time or, rather, the inverse temperature. Furthermore, its value also distinguishes between the non-relativistic ($x \gg 3$), semi-relativistic ($x \approx 3$), and highly relativistic ($x \ll 3$) regions. Note that the velocity in terms of the variable $x$ is given by $v = \sqrt{\frac{3}{x}}$.

Note that, in order to express the coupling constant $G$ in terms of the DM-nucleon cross-section as used in Ref.~\cite{Riedel:2012ur}, one must take into account that a DM particle detected in an experiment performed ``today'' (\emph{i.e.}\ when the Universe is about $13.8$~Gyrs old) is non-relativistic, and a typical value of its velocity \emph{today} would be $v_0 \sim 10^{-3}$~\cite{Smith:2006ym,Savage:2006qr}. One can thus estimate:
\begin{equation}
 \langle \sann v_0 \rangle_{\rm approx} \sim \sann v_0 \gtrsim \frac{2}{9} \ \frac{\qed^2}{16 \pi^2} \scat v_0,
 \label{eq:sigma-estimate}
\end{equation}
which, using $x\gg3$ in \eqref{eq:inter}, leads to the estimate
\be
\label{G2}
G^2 \sim  v_0\ \frac{\qed^2}{18 \pi} \frac{\scat}{\mc^2}\,.
\ee

\begin{figure}
\centering
\includegraphics[scale=0.45]{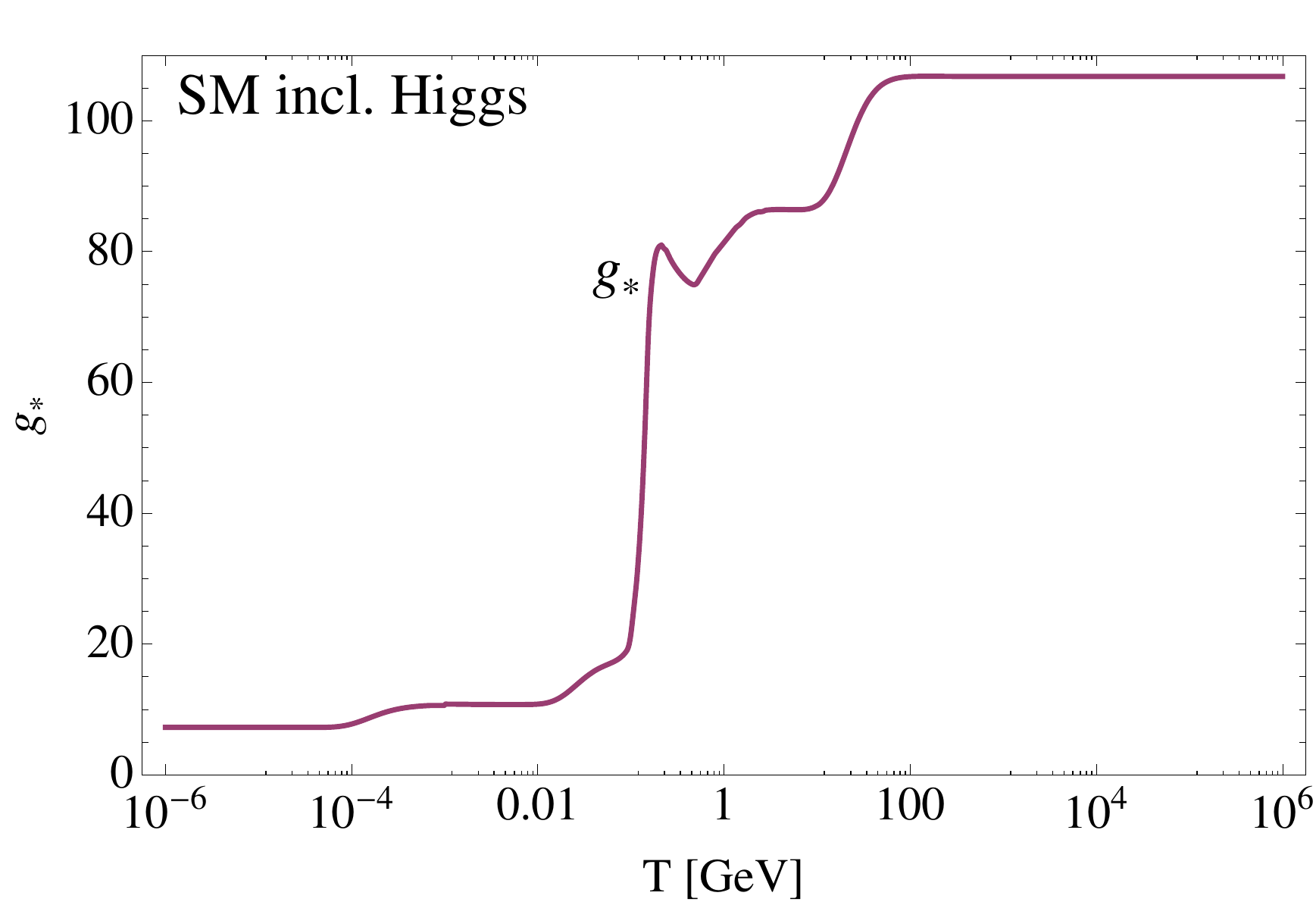}
\caption{\label{fig:gstar}
Effective number $g_*$ of relativistic degrees of freedom for the energy density in the SM, as function of the temperature $T$.
}
\end{figure}

\begin{figure}[t]
\centering
\includegraphics[scale=0.45]{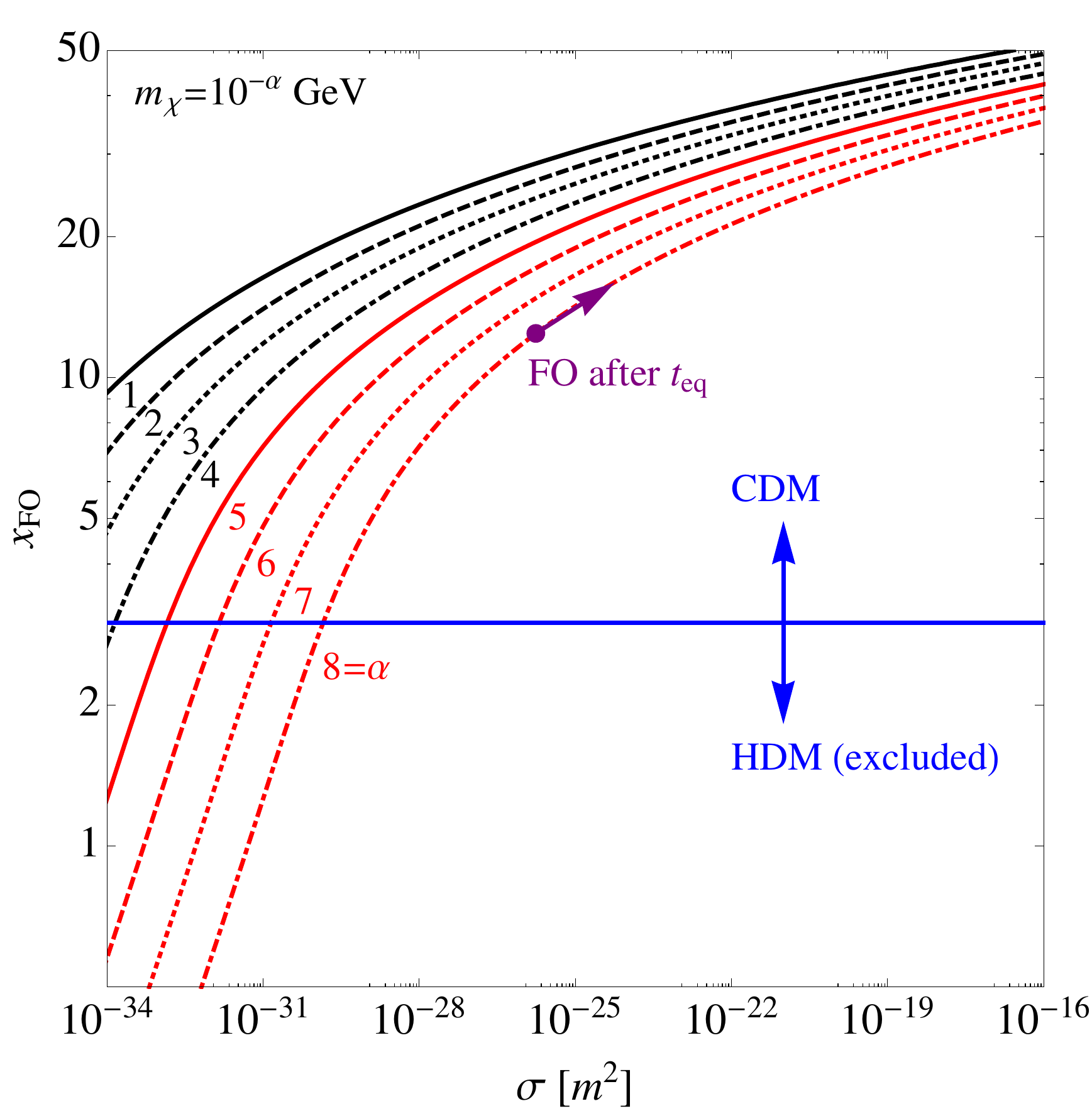}
\caption{\label{fig:xF}
Freeze-out values of the ``time'' $x$, as functions of the cross-section $\scat$. The purple arrow indicates the region where the hypothetical DM candidate would freeze-out after the cosmological matter-radiation equality, which is only important for very small DM masses.
}
\end{figure}

\begin{figure*}
\centering
\hspace{-1cm}
\begin{tabular}{lr}
\includegraphics[width=7.5cm]{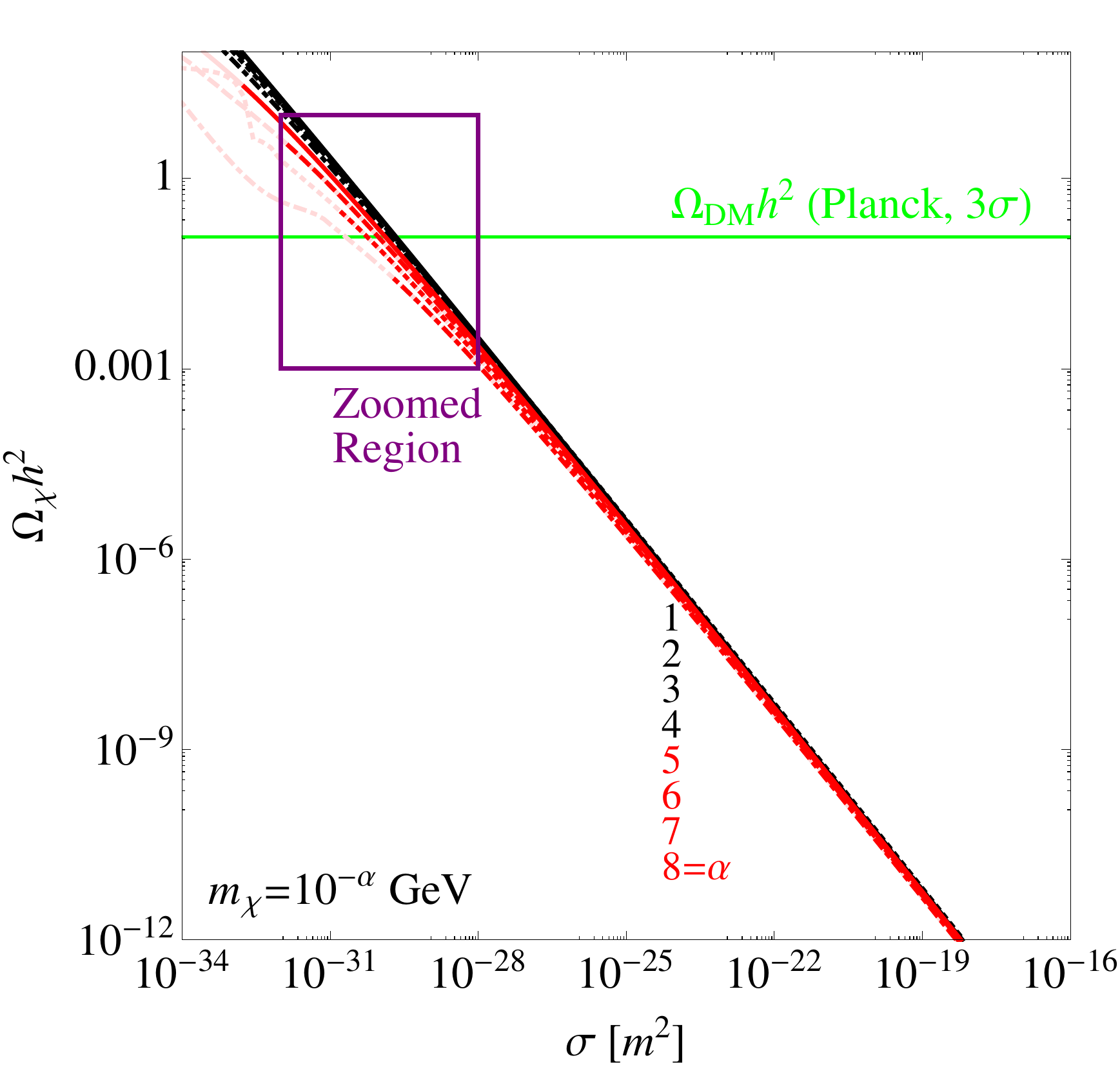} & \includegraphics[width=7.5cm]{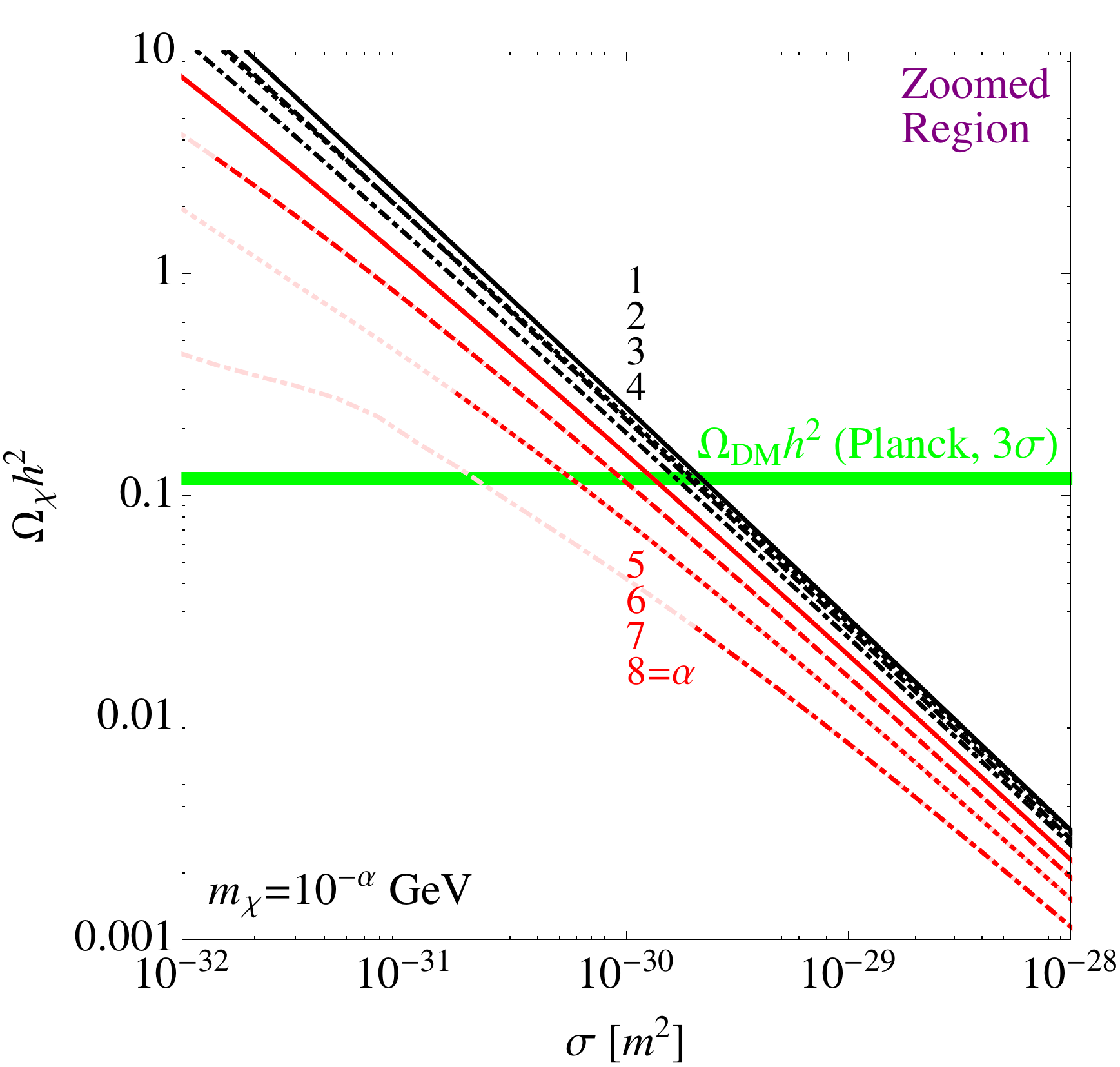}
\end{tabular}
\caption{\label{fig:Omega}
Final DM abundances for different masses of the assumed DM particle, for the whole parameter space considered in Ref.~\cite{Riedel:2012ur} (left panel) and for the most interesting region drawn to a larger scale (right panel). The parts drawn in light colours correspond to hot DM (HDM), which is excluded.
}
\end{figure*}

The first step in the actual calculation is to compute the freeze-out temperature. This determines the final abundance in particular in the case of the cold (non-relativistic) freeze-out, due to the exponential suppression of the number density, $n_\chi \propto e^{- m_\chi / T}$. In order to do this, one needs to equate the interaction rate $n_{\chi, \rm eq} \langle \sann v \rangle_{\rm approx}$ of the DM candidate $\chi$ in thermal equilibrium with the expansion rate $H$ of the Universe,
\begin{equation}
 n_{\chi, \rm eq} \langle \sann v \rangle_{\rm approx} = H.
 \label{eq:FO-condition}
\end{equation}
Due to the relatively large cross-section on quarks necessary for a particle to be detected in a set-up as proposed in Ref.~\cite{Riedel:2012ur}, it is \emph{unavoidable} for the DM particle to be in thermal equilibrium with the photons (or, more precisely, the whole thermal plasma) in the early Universe. To give some more technical details, the expansion rate (Hubble function) of the Universe is in this early and radiation-dominated era given by
\begin{equation}
 H = \frac{1}{2 t},
 \label{eq:Hubble}
\end{equation}
with the time (the age of the Universe) $t$ being related to the temperature $T$ of the Universe (\emph{i.e.}\ the temperature of the thermal plasma) by
\begin{equation}
 t T^2 = \frac{0.301 M_P}{\sqrt{g_*(T)}}.
 \label{eq:t-T-relation}
\end{equation}
Here, $M_P = 1.22 \cdot 10^{19}~{\rm GeV}$ is the Planck mass and $g_*(T)$ is the effective number of relativistic degrees of freedom. This number essentially sums up all the spin and colour degrees of freedom of all particles which are relativistic at a given temperature $T$. For the SM particle content, it is displayed in FIG.~\ref{fig:gstar}. If there is unknown physics beyond the SM, \emph{i.e.}\ many more particles which have non-negligible interactions strengths and can hence be produced in the early Universe, then this function would need to be modified. However, this modification will not be very significant unless very many new particles are introduced. Although we will use the full $g_*(T)$ for the purposes of producing the plots in FIGs.~1 {\bf c}, \ref{fig:gstar}--\ref{fig:Omega}, it is worth noting that for the freeze-out temperatures $T_{\rm FO}<m_\chi/3$ of our eventual choice of DM particle, the effective number of degrees of freedom is to good approximation just $g_*\approx7.25$, corresponding to the fact that only photons and neutrinos can make a contribution.

The equilibrium number density of the DM candidate $\chi$ is (for scalars) given by the ordinary Bose-Einstein distribution,
\begin{equation}
 n_{\chi, \rm eq} = \frac{1}{(2 \pi)^3} \int%\limits_{p=0}^\infty 
 \frac{d^3 p}{e^{E/T}-1},
 \label{eq:n_eq}
\end{equation}
where $E = \sqrt{m_\chi^2 + p^2}$ is the total energy of a particle with mass $m_\chi$ and momentum $p$. 

The inverse freeze-out temperature $x_{\rm FO}$ is obtained by numerically solving Eq.~\eqref{eq:FO-condition}. The result is displayed, for different masses $m_\chi$ in FIG.~\ref{fig:xF}, as a function of $\scat$. We have indicated the region where the DM would be hot (\emph{i.e.}\ highly relativistic at freeze-out), which is excluded because this scenario would not lead to a successful formation of structures in the Universe~\cite{Abazajian:2004zh,dePutter:2012sh}. Furthermore, in the region right of the purple point (where $m_\chi = 10^{-8}$), the freeze-out would happen too late, \emph{i.e.}\ after the time where the energy densities of matter and radiation must have been equal according to observations (corresponding to the temperature $T_{\rm eq}\approx0.8$~eV), significantly affecting the measured fluctuations in the CMB.~\footnote{This holds unless the standard history of the Universe would be considerably altered, which is possible but appears unlikely from the conservative point of view.} However, as we will see later on, that part of the curve is in any case excluded. This problem does not appear for larger masses $m_\chi$, which is why there are no corresponding markings in the plot.

Writing $p=m_\chi y$ in \eqref{eq:n_eq}, where $y$ is a dimensionless integration variable, and using the fact that we want $x=m_\chi/T>3$ in order for DM not to be too hot, the integral may be approximated to better than 6\% by
\begin{equation}
\label{N-approx}
n_{\chi, \rm eq}\approx {m_\chi^3\over2\pi^2}\,{\rm e}^{-x}
\sqrt{\pi\over2x^3}\left(1+{15\over8x}\right)\,.
\end{equation}
Using $g_*\approx7.25$ and combining the expressions above we thus have: 
\be
\label{sigma-approx}
\scat\approx {0.9613\cdot10^{-37}\ {\rm m}^2
\over \mc\ [{\rm GeV}]} {(1+\xfo)\,\xfo^{5/2}\,{\rm e}^{\xfo}\over(12+12\xfo+5\xfo^2+4\xfo^3)(8\xfo+15)}\,.
\ee
Although it is slightly counter-intuitive to instead express $\sigma$ as a function of $m_\chi$ and $\xfo$ in this way, we thus get a closed expression which agrees with the plots of FIG.~\ref{fig:xF}; for example it accounts for the $\sigma\propto1/m_\chi$-dependence clearly visible in the plots.

Using the standard techniques, one can then expand the cross-section as $\sann v = a + b \langle v^2 \rangle + \mathcal{O}(v^4)$, where $\langle v^2 \rangle = 6/x$, leading to
\begin{equation}
 a \sim \frac{G^2 m_\chi^2}{4 \pi},\ \ b \sim \frac{a}{24},
 \label{eq:a_b}
\end{equation}
where $G^2$ was reported in Eq.~\eqref{G2}. The standard formula for the final DM abundance is given by~\cite{Drees:2009bi}:
\begin{equation}
 \Omega_\chi h^2 = \frac{8.5\cdot 10^{-11}\ {\rm GeV}^{-2}\ x_{\rm FO}}{\sqrt{g_*(x_{\rm FO})} (a + 3b / x_{\rm FO})} \sim \frac{4.4\cdot 10^{-35} x_{\rm FO}}{\sqrt{g_*(x_{\rm FO})} [1 + 1 / (8 x_{\rm FO})]\cdot \scat [{\rm m}^2] v_0},
 \label{eq:abundance}
\end{equation}
where the cross-section is measured in square metres. Note that, as pointed out in Ref.~\cite{Drees:2009bi}, using in Eq.~\eqref{eq:abundance} the value of $x_{\rm FO}$ obtained by numerically solving Eq.~\eqref{eq:FO-condition} may lead to errors of something like $10\%$ in the case of non-relativistic freeze-out, so that in the worst case, there could be a further $\mathcal{O}(1)$ factor involved in our result. However, this uncertainty, while present, is less than the uncertainty introduced by using already generic estimates for the cross-sections.

The final result for the abundance is presented in FIG.~\ref{fig:Omega}, where the whole parameter region (masses and cross-sections on nucleons as taken in Fig.~5a of Ref.~\cite{Riedel:2012ur}) is displayed on the left panel, and a blow-up of the most interesting region is shown on the right. As can be seen, the resulting DM abundance can hit the observed value for a ballpark of masses, from $m_\chi = 10^{-1}~{\rm GeV}$ to $10^{-7}~{\rm GeV}$. Smaller masses are excluded even though the correct abundance could be in principle be obtained, because that part of the parameter space would correspond to HDM which is ruled out (or, rather, bound to make up at most about $1\%$ of the total DM in the Universe~\cite{Abazajian:2004zh,dePutter:2012sh}). 

Recalling the cold DM (CDM) constraint $\xfo>3$ and the DM abundance \eqref{odm}, and substituting $g_*\approx7.25$ and the typical value $v_0\sim10^{-3}$ into \eqref{eq:abundance}, we can write compactly that the fraction of DM made up of $\chi$ species is $f={\Omega_\chi/\Omega_{\rm DM}} \approx 1.4\cdot10^{-31}\, {\xfo/\scat [{\rm m}^2]}$. For a fixed fraction $f$, for example $f=1$ corresponding to the case where DM is entirely made up of $\chi$,  the mass $\mc$ is then related to the freeze-out temperature $\xfo>3$ as:
\be
\label{mc-approx}
\mc \approx 7.0\cdot10^{-7}\, f \,{(1+\xfo)\,\xfo^{3/2}\,{\rm e}^{\xfo}\over(12+12\xfo+5\xfo^2+4\xfo^3)(8\xfo+15)}.
\ee
This may be combined with Eq.~\eqref{sigma-approx} to get directly the relevant points in FIG.~\ref{fig:Omega}.

These results may look somewhat surprising at first sight, since for scattering cross-sections around $\scat = 10^{-30}~{\rm m}^2 = 10^{-26}~{\rm cm}^2$, where an abundance in the correct ballpark is generated, one might naively expect a much larger annihilation cross-section which would keep the DM in equilibrium until it is very cold, thereby suppressing its abundance by a huge number. However, because of the DM particles being so light, the strong suppression of the annihilation cross-section (due to it necessarily being a loop process in this case) saves our DM candidate from that fate.\\

\begin{figure*}
\centering
\hspace{-1cm}
\begin{tabular}{lr}
\includegraphics[scale=0.5]{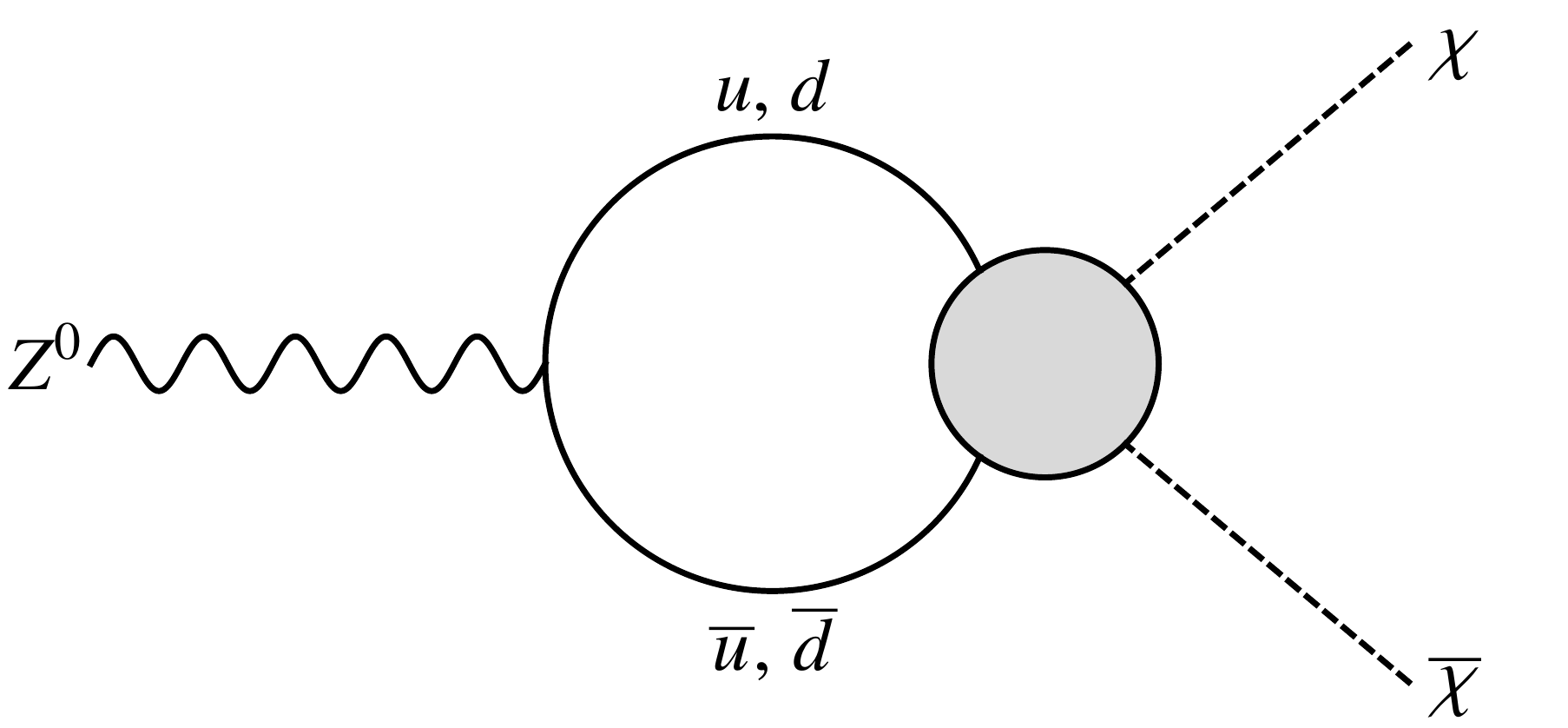} & \includegraphics[scale=0.5]{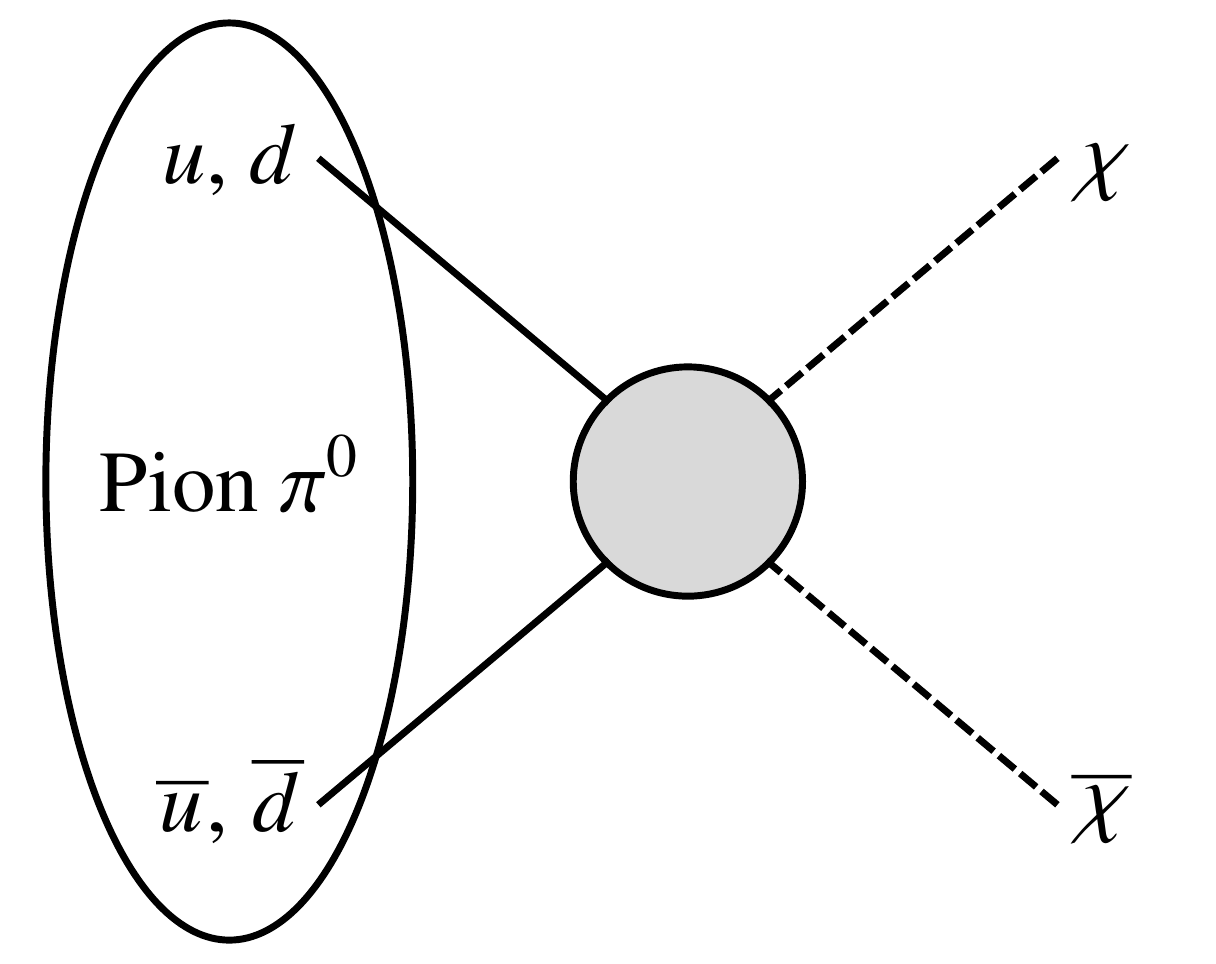}
\end{tabular}
\caption{\label{fig:Decays}
Diagrams for the potential invisible decays of the $Z$-boson (left) and of the neutral pion (right).
}
\end{figure*}

\paragraph{Particle physics constraints}

While the calculation up to this point looks in fact quite good, we should mention that there could be potentially dangerous bounds, because after all our DM candidate does couple quite strongly to quarks.

As we had mentioned, one could expect a potentially strong bound from hadron colliders, such as the LHC. While the most natural choice to search for a $\chi$ particle may seem to be a proton-antiproton collider such as the Tevatron, the most natural reaction $p \bar{p} \to \chi \bar{\chi}$ would probably be invisible since the $\chi$ would just remain on the beam direction due to their small masses and the associated large boost factors. However, if one of the quarks inside the baryons radiated off a photon before the annihilation process, this could lead to a classic signature of one single photon plus missing energy.~\footnote{As already mentioned, $\chi$ would not necessarily show up as missing energy. For low energies, it would be shielded by the detector material, but for high energies it could deposit energy in a calorimeter. However such a signal would be generated primarily from a $\chi$ particle embedded in a jet, through multiple $t$ channel exchanges, and thus easily confused with other particles. (This follows from the high energy $s$-channel suppression we are about to discuss.)} Thus, the most stringent bounds come from experiments performed at hadron colliders which have searched for such a signature, for example CDF~\cite{Aaltonen:2008hh}, D0~\cite{Abazov:2008kp}, or CMS~\cite{CMS:2011tva}.~\footnote{Note that bounds from lepton colliders such as LEP do not play any role as long as our DM candidate does not couple directly to leptons. Thus the classic limits from detectors like DELPHI~\cite{Abdallah:2003np,Abdallah:2008aa} may not apply in this case.} However, depending on the true ultraviolet completion of the effective vertex in FIGs.~1~{\bf a,b} (main text), this may not be a problem. Imagine the existence of another scalar particle $\xi$ with mass $m_\xi$ which couples with strength $g_q$  to quark/antiquark pairs $q \bar{q}$, and with the dimensionful 3-point coupling $f_\chi$ to DM/anti-DM pairs $\chi \bar{\chi}$. If it holds that $m_\chi \ll m_\xi \ll E_{\rm collider}$, \emph{i.e.}\ the new particle is much heavier than $\chi$ but has a mass much smaller than typical collider energies, the interaction between quarks and $\chi$'s would be an effective four-point coupling  for all low energy purposes as already discussed, but fundamental at collider level. If furthermore $g_q$ is very small, the $\xi$ would not show up in any electroweak precision data.

The question remains, however, if this could possibly lead to a large enough scattering cross-section for $\chi$. This point can be easily answered. Taking the collider limits to amount to roughly $\sigma_{\rm ann.+\gamma} < 1$~fb for a centre-of-mass energy of $\sqrt{s} \sim 100$~GeV (which is already conservative), then the cross-section can be estimated on dimensional grounds as $\sigma_{\rm ann.+\gamma} \sim \frac{g_q^2 f_\chi^2}{s^2} \qed$, where the additional suppression factor comes from the photon being radiated off. Taking $\qed \sim 10^{-2}$, one can thus estimate $g_q^2 (f_\chi/{\rm GeV})^2 < 0.1$. At low energies, in turn, the scattering cross-section would be $\sigma_{\rm scatt.} \sim \frac{g_q^2 f_\chi^2}{m_\xi^4}$. The scattering cross-section needed for the correct abundance is roughly $\sigma \sim 10^{-30}~{\rm m}^2$, \emph{cf.}\ FIG.~\ref{fig:Omega}, which has to be divided by another factor of roughly $10$ to translate it from nucleons into quarks. Using the collider bound, one can estimate $m_\xi \lesssim 0.3$~GeV, which is still much larger than the required masses for $\chi$. Thus, a sufficiently light $\xi$ can compensate for the small coupling $g_q$ such that the scattering cross-section can be large at low energies but suppressed at collider level. This argumentation is quite generic and would make the strong limits from colliders much less problematic (if not completely harmless).

Strong bounds may also originate from the invisible decay width of the SM $Z$-boson, which could in principle decay as $Z^0 \to \chi \overline{\chi}$ by a 1-loop diagram with $u$- and $d$-quarks intermediate states (\emph{cf.}\ left panel of FIG.~\ref{fig:Decays}), and from the invisible decay of a neutral pion, $\pi^0 \to \chi \overline{\chi}$, again involving a $u$- and/or $d$-loop (\emph{cf.}\ right panel of FIG.~\ref{fig:Decays}). This might lead to a problem, because the invisible decay width of the $Z$-boson is well measured, $\Gamma_Z^{\rm inv} = 499.0 \pm 1.5$~MeV~\cite{Beringer:1900zz}, and it agrees with the SM prediction so that any additional contribution could only modify it by an amount of the order of the uncertainty $1.5$~MeV of the measurement. There is no actual measurement on the invisible pion decay width available, but a bound can be estimated from the branching ratio of $2.7\cdot 10^{-7}$ (at $90\%$~C.L.) of $\pi^0$~\cite{Beringer:1900zz} decaying into neutrinos which results into a tiny invisible decay width of $\Gamma_{\pi^0}^{\rm inv} < 2\cdot 10^{-12}$~MeV. Indeed, such very model-independent bounds can exist for light DM candidates (see, \emph{e.g.}, Refs.~\cite{King:2012wg,DiBari:2013dna} for concrete examples).

However, for the scalar DM case, both these processes in fact have vanishing amplitudes. This is relatively easy to understand. In the case of the $Z$-boson, since the $\chi$ particle has no charge or hypercharge, there is no gauge-invariant effective coupling to mediate $Z^0 \to \chi \overline{\chi}$. Although gauge symmetry is spontaneously broken,  the one-loop diagram illustrated above is insensitive to this. In fact by this argumentation one sees that the simplest coupling involving $Z$ and $\chi$ is via the dimension six operator $\sim B^2_{\mu\nu} \overline{\chi}\chi$, where $B_{\mu\nu}$ is the $U(1)_Y$ field strength. This allows for processes such as $Z^0 \to \chi \overline{\chi}\gamma$, but they will be suppressed by $\qed$. For the pion decay, in turn, one has to take into account that the pion is in fact a pseudo-scalar (\emph{i.e.}\ the corresponding field changes its sign under parity transformations), but the $s$-wave final state containing two scalars will always be of positive parity. Thus, the amplitude is zero as long as no parity breaking interactions are assumed, since the kinematics of the situation always enforce back-to-back emission of the final states and their spinlessness makes it impossible to compensate for that by orbital angular momentum. Again, the simplest process that allows the pion to decay to $\chi$ particles must involve also radiating a photon and thus is suppressed by $\qed$. In spite of these arguments, we have calculated both processes explicitly (using a scalar vertex $\mathcal{L}_{\rm eff} = \tilde \xi |\chi|^2 \overline{q} q$ for the $Z$-boson diagram and the chiral Lagrangian in combination with Eq.~\eqref{eq:eff_nucleon_int} for the pion decay diagram), and we can confirm that the amplitudes are indeed zero. While these bounds would certainly be strong for a number of possible DM candidates, a scalar $\chi$ evades them completely. Again, this type of particle seems to save itself from the most dangerous bounds. However, this situation could be very different for other possible DM candidates, such as fermionic particles. Furthermore, fermionic DM would necessarily obey the so-called Tremaine--Gunn bound~\cite{Tremaine:1979we}, which is essentially based on the Fermi pressure of fermions. Recent analyses~\cite{Boyarsky:2008ju} of dwarf satellite galaxies show that, using the limit of a degenerate Fermi gas, a lower bound of $m_{\rm DM}^{\rm fermionic}>0.41$~keV is derived on the mass of a fermionic DM particle. This would indeed cut significantly into our parameter space. Note that, in our case, it is not unthinkable that only a certain fraction of the DM, say $p=1\%$, is made up of the $\chi$ particles studied in this work. In that case, the mass bound does become marginally lower by a factor of $p^{1/12}$~\cite{Boyarsky:2008ju}, \emph{e.g.}, $m_{\rm DM}^{\rm fermionic}>0.28$~keV for $p=1\%$.\\

\paragraph{The annihilation signal}

The annihilation of two DM particles into two photons, \emph{cf.}\ FIG.~1~{\bf b} (main text), is the only possible annihilation channel whenever the two initial state particles are non-relativistic. Thus in regions where a lot of DM particles accumulate, such as the Galactic centre, the same process will be active and could possibly lead to an observable signal.

The differential flux of photons per area and time, stemming from the annihilation of two DM particles, can  easily be calculated~ \cite{Bergstrom:1997fj,Bergstrom:2004cy}:
\begin{equation}
 \frac{d \Phi_\gamma}{d E_\gamma} \simeq 9.3\cdot 10^{-3} \left( \frac{\rm GeV}{m_\chi} \right)^2 \frac{d N_\gamma}{d E_\gamma} \left( \frac{\sigma_{\rm ann}}{10^{-32} {\rm m}^3 {\rm s}^{-1}} \right)\ \langle J_{\rm GC} \rangle_{\Delta \Omega}\Delta \Omega\ {\rm m}^{-2} {\rm s}^{-1}.
 \label{eq:sig_1}
\end{equation}
Since the broadening of the differential spectrum is small and not of any relevance if we are only interested in the total photon flux, we can take $d N_\gamma / d E_\gamma = 2 \delta (E_\gamma - m_\chi)$, where the factor of $2$ originates from the fact that two photons are produced per annihilation process. We calculate the expected flux from the Galactic centre, for the sake of an example, for now assuming that the signal could reach us without being perturbed (as we will see later on, our Galaxy is in fact opaque for part of the photon spectrum). For a Navarro-Frenk-White (NFW) profile~\cite{Navarro:1995iw} with the parameters $(\alpha, \beta, \gamma) = (1.0, 3.0, 1.0)$~\footnote{This $\alpha$ is not to be confused with the parameter we use to parametrise the DM mass according to $m_\chi = 10^{- \alpha}$~GeV.} and a scale radius of $r_S = 20$~kpc, one expects $\langle J_{\rm GC} \rangle_{\Delta \Omega}\Delta \Omega = 0.13$~sr for the line-of-sight integral, where $\Delta \Omega = 10^{-5}~{\rm sr}$.~\footnote{Note that the angular resolution $\Delta \Omega$ is an experimental parameter, which varies from telescope to telescope, and thus implicitly with energy. However, there are instruments available which achieve $\Delta \Omega = 10^{-5}$~sr, so that we consider this as a good choice for an example calculation.} Thus, the integrated total flux at an energy of $E_\gamma \simeq m_\chi$ is given by:
\begin{equation}
\Phi_{\rm tot} \simeq 2.4\cdot 10^{-3} \left( \frac{\rm GeV}{m_\chi} \right)^2 \left( \frac{\sigma_{\rm ann}}{10^{-32} {\rm m}^3 {\rm s}^{-1}} \right)\ {\rm m}^{-2} {\rm s}^{-1}.
 \label{eq:sig_2}
\end{equation}
As can be seen already from this formula, the smallness of $m_\chi$ leads to a huge photon flux. This can be understood intuitively: since the mass of the DM candidate under consideration is comparatively small, its number density must be quite high to compensate for the small mass so that the correct DM abundance can be met. Since there are no kinematical restrictions associated with the annihilation into two photons (the only restrictions could come from angular momentum related issues, but these are not present here), this rate cannot depend strongly on the initial state masses, apart from them being the only dimensionful quantities involved. Thus, the large number density translates into a high photon rate, since the fact that two DM particles have to meet to annihilate yields to a proportionality of the signal rate to a square of the DM number density.

The resulting fluxes are plotted in FIG.~1~{\bf c} (main text). Indeed, the fluxes turn out to be very large. So large, in fact, that they would exceed the known bounds by orders of magnitude, if taken at face value (\emph{i.e.}\ if no further subtleties such as strong atomic lines are considered). This would also be true for annihilation in regions other than the Galactic centre. As we will show, we have as example computed the expected rate for a DM mass of around $3.56$~keV in order to see whether we could reproduce the recently reported X-ray signal from galaxy clusters~\cite{Bulbul:2014sua,Boyarsky:2014jta}, and indeed our DM candidate would, due to the large annihilation cross-sections related to the large direct detection cross-sections, exceed the observed signal strength by several orders of magnitude. Similar results would be obtained when computing the signal for some other regions in the parameter space.

Accordingly, the natural reaction one could (and probably should) have is to discard the DM candidate particle discussed in this article, because its large annihilation signal would already have been seen for sure.~\footnote{Even though the resulting peak can be expected to be quite narrow, it would seem quite unrealistic that such a giant signal could possible have been missed due to insufficient energy resolution.} But would this conclusion be correct? As it turns out, it would not! The simple reason is that, first of all, not all energy ranges relevant here have been thoroughly investigated by observations, as some of them were considered to be ``uninteresting'' from an astrophysical point of view, but furthermore a galaxy can also be quite opaque to certain wavelengths, so that the signal, even if present, would not necessarily reach us. As we will see, this leaves us with an unconstrained window in the parameter space. Furthermore, even for the regions where we exceed the bounds, there are many subtleties involved with detecting a line signal, ranging from potentially strong backgrounds by neighbouring atomic lines to the modelling of the continuum background. We are aware of these subtleties but we cannot discuss them here, since in particular many of them are very specific to certain energy ranges, while our global analysis spans over many orders of magnitude in energy. However, to make clear that one would need to investigate certain regions in greater detail to be absolutely sure that our DM candidate could not hide in there, we marked the regions threatened by astrophysics as ``disfavoured'' rather than ``excluded'' in FIG.~1~{\bf c} (main text), and we indicate that by using only a light gray background colour.

Before discussing the tentative bounds, we need to be clear which photon energy range we are talking about. Since $E_\gamma \simeq m_\chi$, any strong constraint on $m_\chi$ will directly translate into a constraint on $E_\gamma$. As we have already seen, \emph{cf.}\ FIG.~\ref{fig:Omega} (in particular for $m_\chi = 10^{-8}$~GeV$= 10$~eV), for too small masses of $\chi$ we are in fact hitting the HDM region, which would conflict with cosmological structure formation. In addition recall that the freeze-out temperature $T_{\rm FO}$ has to be greater than the equality temperature $T_{\rm eq}$.  This sets a lower limit $m_\chi > 3T_{\rm FO}>3T_{\rm eq}= 2.4$~eV for all cross-sections under consideration, which results in the upper dark gray exclusion region in FIG.~1~{\bf c} (main text). If $m_\chi > m_\pi$, in turn, the suppression of the annihilation cross-section, \emph{cf.}\ FIGs.~1~{\bf a,b} (main text), would not work anymore, thereby completely destroying any abundance of $\chi$, which translates into the lower dark gray region in the plot. The region in between these boundaries can be constrained by observations.

The first question to answer is where to look for signals. Dwarf spheroidal galaxies are generally considered to be very good places to search for DM decay or annihilation lines, as they have a very high mass to light ($M/L$) ratio and are therefore thought to contain a particularly high fraction of DM. Clusters of galaxies are also good as they have a high $M/L$ ratio, too, except that they are usually strong sources of bremstrahlung X-ray emission which forms a high background against which to search for the X-ray signal lines. Some other searches have looked at various angles in our own Galaxy. DM decay or annihilation lines would be expected to be stronger nearer to the Galactic centre, due to the accumulation of DM, and so any lines which did not vary appreciably around the sky are probably not decay or annihilation lines.

Which observations are relevant for us? Let us start by the upper energy boundary and work (roughly) down in energy. There have been a number of searches for decay lines in the Fermi data. In particular Ref.~\cite{Abdo:2010ex} presented observations of dwarf spheroidal galaxies. Above $100$~MeV they find no detections and derive upper limits on any line of approximately a few times $10^{-9} {\rm cm}^{-2} {\rm s}^{-1}$ (per $10^{-5}$~sr), which is below the signal expected from our region of interest. In Ref.~\cite{Boyarsky:2007ge}, a search for lines in data taken with the high resolution spectrometer, SPI, on the INTEGRAL $\gamma$-ray observatory has been presented. The search is performed in the energy range from $40$~keV to $14$~MeV, in blank sky data at various distances from our own Galactic centre. They find a number of lines, many of which are identified as instrumental lines, but some of which are unidentified. However, none of these lines varies enough to be considered a likely DM line. They place upper limits on the DM origin of each line, and these limits typically lie in the range $10^{-7}$ to $10^{-8} {\rm cm}^{-2} {\rm s}^{-1}$. Again, a signal from our DM candidate in that mass range would completely overshoot these bounds, if taken at face value.

In the keV region, a vast variety of bounds exist~\cite{Dolgov:2000ew,Abazajian:2001vt,Boyarsky:2005us,Watson:2006qb,Abazajian:2006jc,Boyarsky:2006fg,RiemerSorensen:2006fh,Abazajian:2006yn,Boyarsky:2006ag,Boyarsky:2007ay,Boyarsky:2007ge,Loewenstein:2008yi,Watson:2011dw,Loewenstein:2012px}. However, since these bounds are typically interpreted in terms of keV sterile neutrino DM (see, \emph{e.g.}, Refs.~\cite{Kusenko:2009up,Merle:2013gea} for reviews), the corresponding fluxes, which on top of that come from many different observations of various (dwarf) satellite galaxies, are usually given in terms of the so-called \emph{active-sterile mixing angle} $\theta$. For our purpose, it is most reasonable to translate these bounds into event rates, in order to compare them to our cross-sections. This job is made easy by recently proposed fit formulas~\cite{Merle:2013ibc} to the combination of the bounds reported above.

To perform the comparison, it is easiest to look at event rates instead of fluxes. For sterile neutrino decays, the photon rate per second in a given volume $V$ can be estimated as $V n_{\rm DM} \Gamma_\gamma$, where
\begin{equation}
 \Gamma_\gamma \simeq 1.38\cdot 10^{-29} {\rm s}^{-1} \left( \frac{\sin^2 (2\theta)}{10^{-7}} \right) \left( \frac{m_s}{\rm keV} \right)^5
 \label{eq:sig_3}
\end{equation}
is the decay rate of a sterile neutrino with mass $m_s$~\cite{Bulbul:2014sua}. For the corresponding annihilation rate of our DM candidate, the equivalent event rate is $V n_{\rm DM}^2 \sigma_{\rm ann}$, where $\sigma_{\rm ann}$ is given in Eq.~\eqref{eq:sigma-estimate_3} and the square on the number density arises from the fact that two DM particles have to meet in order for the annihilation to take place. Since the event rate from our candidate has to be smaller than the bound, we can easily derive
\begin{equation}
 \sigma \leq 9.84\cdot 10^{-27} {\rm m}^2\cdot m_\chi[{\rm GeV}]\cdot \theta^2\left( \frac{m_s}{\rm keV} \right)^5,
 \label{eq:sig_4}
\end{equation}
where we have used the fact that $\theta$ is small, and we have expressed the DM number density by the local DM energy density, $n_{\rm DM} = \rho_\chi/m_\chi$ with $\rho_\chi = 0.4~{\rm GeV}/{\rm cm}^3$. Note that, for the comparison, we need to set $m_s = 2 m_\chi$, due to the sterile neutrino signal arising from a 2-body decay $\nu_s \to \nu \gamma$, whereas our signal would arise from an annihilation process.

Using the conservative bounds from Ref.~\cite{Merle:2013ibc}, the upper bound on the cross-section $\sigma$ turns out to be around $10^{-38}~{\rm m}^2$, for the whole mass range from $m_\chi = 0.25$--$25$~keV. Thus, a signal from our DM candidate would probably not have been missed in this mass range, so that it is strongly disfavoured for our purpose as well. Alternatively, we could try to reproduce the recently reported $3.56$~keV X-ray line signal~\cite{Bulbul:2014sua,Boyarsky:2014jta}. The derived mixing angle of $\sin^2 (2\theta) = 7\cdot 10^{-11}$ would again require a cross-section of $\sigma \approx 10^{-38}~{\rm m}^2$, which is off our plot.

We should compare the above non-detections with the typical value of the X-ray background, which provides a lower ball-park sensitivity limit. At 1 keV, Ref.~\cite{Lumb:2002} gives the background as $1.1\cdot 10^{-4}$ photons per ${\rm cm}^2 {\rm s}\ {\rm keV}$ (per $10^{-5}$~sr). Taking a conservative energy resolution of $10\%$, a $3\sigma$ limit on the flux would be approximately $3.3\cdot 10^{-5} {\rm cm}^{-2} {\rm s}^{-1}$. Using typical broad band diffuse X-ray background measurements (\emph{e.g.}~\cite{Gilli:2003bm}), we obtain $3\sigma$ limits which are factors of about 10 below those listed in Fig.~2 of~\cite{Boyarsky:2006fg}. That is the direction that one would expect so the observations of the background are merely a weak consistency check, and again, a signal of a DM candidate like ours would have been highly visible.

However, in the range of approximately $10$~eV ($13.6$~eV, to be precise) to about $100$--$200$~eV, observations are severely limited due to the absorption by neutral hydrogen in our own Galaxy. Thus, in that energy range our Galaxy is in fact opaque and a photon signal from the Galactic centre cannot be expected to reach us. So, indeed, this mass/energy range at the moment comprises an astrophysical window, in which our DM candidate could live, \emph{cf.}\ the white band in FIG.~1~{\bf c} (main text). As indicated by the purple band in the plot, there is a surviving and distinctive region in which our DM candidate could live, even when putting all known constraints together. This narrows us down so far that we can characterise the properties of the DM particle presented here to be a scalar particle $\chi$ with
\begin{equation}
m_\chi \approx 100~{\rm eV}\ \ \ {\rm and}\ \ \ \sigma \approx 5\cdot 10^{-31}~{\rm m}^2.
 \label{eq:DM-properties}
\end{equation}
This is the parameter region which should be scrutinised and where, ultimately, experimentalists should search if they plan to probe our proposal.

A cautionary note at the end: The predictions given in the current paper relate to the surface brightness of the expected emission. Surface brightness, at least in the local universe, is independent of the distance. Thus, to first order, the surface brightness from DM decay radiation around a line of sight through our Galactic Centre (GC) should be similar to that through the centre of a similar nearby galaxy. However, many of the nearby dwarf spheroidal galaxies, which are promising targets for DM detection, have angular sizes less than a degree, making them, at best, marginally resolved in Fermi observations ($>1$~GeV) and thus Fermi search papers give total integrated fluxes (\emph{e.g.}~\cite{Abdo:2010ex}). A proper comparison with theory then requires that the emission from an assumed DM density profile is integrated for comparison with observation. The resulting integrated flux will thus be less than if the central surface brightness flux prevailed over the whole of the resolution element. However, for the nearby dwarf spheroidals, where the angular scale size of the galaxy is not much smaller than the resolution size of the instrument, the difference in integrated fluxes is unlikely to be more than a factor of 10, which is pretty marginal for the regions we can rule out, since a potential signal would overshoot the bounds by much more than that.

\subsection{Acceleration of a spherical test particle}
As described in the main text, a $\chi$ particle at a typical velocity has a de Broglie wavelength large compared with the inter-nuclear separation of normal matter, and so the overall effect of multiple scattering events is well described as an interaction with an effective potential; cold neutrons interact with normal matter in a similar way~\cite{sears1989neutron}. We use partial waves to treat a spherically-symmetric test particle which we describe as a finite potential well with radius given by the size of the particle and a depth chosen to match the scattering cross-section \eqref{eq:DM-properties} at low energy.

The wavefunction for a $\chi$ particle incident from a distant source is well approximated by a plane wave, and the total wavefunction after scattering by a localised particle is
\begin{equation}
  \label{eq:2}
  \psi\propto e^{ikz}+f(\theta)\frac{e^{ikr}}{r},
\end{equation}
where $k$ is the wavenumber of the incident particle, and $r$, $\theta$ are the radial coordinates relative to the scatterer and the $z$ axis, respectively. From scattering theory, we identify the differential cross-section
\begin{equation}
  \label{eq:3}
  \frac{d\sigmatotal}{d\Omega}=\lVert f(\theta)\rVert^2
\end{equation}
which, using partial waves, we can express as
\begin{equation}
  \label{eq:4}
f(\theta)=\frac{1}{k}\sum_{l=0}^\infty (2l+1)e^{i\delta_l}\sin\delta_l P_l(\cos\theta)=\frac{1}{k}\sum_{l=0}^\infty c_l P_l(\cos\theta),
\end{equation}
where $c_l\equiv (2l+1)e^{i\delta_l}\sin\delta_l$, $\delta_l$ is the phase-shift for angular momentum $l$, and $P_l$ are the Legendre polynomials.
For a pressure $P$, the force on the particle is the flux multiplied by the cross-section, less the recoil at angle $\theta$:
\begin{eqnarray}
F &=& P \int_\Omega \frac{d\sigmatotal}{d\Omega}\left(1-\cos\theta\right)d\Omega \nonumber\\
&=&P\sigmatotal - P~2\pi \int_0^\pi\cos\theta \lVert f(\theta) \rVert^2 \sin\theta d\theta.
\end{eqnarray}
Expanding by using Eq.~\eqref{eq:4}, we find
\begin{eqnarray}
\frac{F}{P}&=&\sigmatotal-\frac{2\pi}{k^2}\int_0^\pi\sin\theta\cos\theta \sum_{l=0}^\infty\sum_{m=0}^\infty c_l^* c_m P_l(\cos\theta)P_m(\cos\theta)d\theta \nonumber\\
&=&\sigmatotal-\frac{2\pi}{k^2}\sum_{l=0}^\infty\sum_{m=0}^\infty c_l^*c_m\int_0^\pi\sin\theta\cos\theta P_l(\cos\theta)P_m(\cos\theta)d\theta \nonumber\\
&=&\sigmatotal-\frac{2\pi}{k^2}\sum_{l=0}^\infty\sum_{m=0}^\infty c_l^*c_m\int_{-1}^{+1}x P_l(x)P_m(x)dx \nonumber\\
&=&\sigmatotal-\frac{2\pi}{k^2}\left(\sum_{l=0}^\infty c_l^* c_{l+1}\frac{2l+2}{(2l+2)^2-1} + \sum_{l=1}^\infty c_l^* c_{l-1}\frac{2l}{(2l)^2-1}\right),
\end{eqnarray}
where we have used
\begin{equation}
  \label{eq:5}
  \int_{-1}^{+1}xP_l(x)P_m(x)dx=\left(\delta_{l,m+1} + \delta_{m,l+1}\right)\frac{l+m+1}{(l+m+1)^2-1}.
\end{equation}
Here, $\delta_{a,b}$ is the Kronecker delta. Relabelling with $l'=l-1$, the second summation becomes the complex conjugate of the first, and
\begin{equation}
  \label{eq:6}
\frac{F}{P}=\sigmatotal-\frac{4\pi}{k^2}\sum_{l=0}^\infty \mathcal{R}\left[c_l^*c_{l+1}\right]\frac{2l+2}{(2l+2)^2-1}.
\end{equation}
Using the definitions of $c_l$ to expand $c_l^*c_{l+1}$,
\begin{equation}
  \label{eq:7}
c_l^*c_{l+1}=(2l+1)(2l+3)e^{-i\left(\delta_l-\delta_{l+1}\right)}\sin\delta_l\sin\delta_{l+1},  
\end{equation}
we obtain
\begin{equation}
  \label{eq:8}
\frac{F}{P}=\frac{4\pi}{k^2}\sum_{l=0}^\infty\left[ (2l+1)\sin^2\delta_l-\frac{(2l+1)(2l+2)(2l+3)}{(2l+2)^2-1}\cos\left(\delta_l-\delta_{l+1}\right)\sin\delta_l\sin\delta_{l+1}\right]
\end{equation}
where we have used 
\begin{equation}
  \label{eq:9}
\sigmatotal=\frac{4\pi}{k^2}\sum_{l=0}^\infty(2l+1)\sin^2\delta_l.
\end{equation}
This expression is numerically efficient to evaluate.  To accurately describe a $\chi$ particle of wavenumber $k$ scattering from a test particle of radius $r$, we must include at least $kr$ terms.

For a finite spherically-symmetric potential-well of radius $r$ and depth parameterised by a wavenumber $\kappa$~\cite{goldberger1964collision}, the phase-shifts are
\begin{equation}
  \label{eq:10}
  \delta_l=\frac{j_l'(k r)-Dj_l(k r)}{n_l'(k r)-Dn_l(k r)},
\end{equation}
where $D\equiv (K/k)j_l'(Kr)/j_l(Kr)$ and $K=\sqrt{k^2+\kappa^2}$. The functions $j_l$ and $n_l$ are related to the Bessel functions of first- and second-kind:
\begin{equation}
  \label{eq:11}
  j_l(x)=\sqrt{\frac{\pi}{2x}}J_{l+\frac{1}{2}}(x)\text{ and }n_l(x)=\sqrt{\frac{\pi}{2x}}Y_{l+\frac{1}{2}}(x),
\end{equation}
with $j_l'(x)=dj_l(x)/dx$ and $n_l'(x)=dn_l(x)/dx$.

In the low-energy limit, where $k\to 0$ and only $s$-wave ($l=0$) scattering is significant, we have
\begin{equation}
  \label{eq:12}
  \sigmatotal = 4\pi \left[r\left(1-\frac{\tan r \kappa}{r \kappa}\right)\right]^2\approx  \frac{V^2\kappa^4}{4\pi},
\end{equation}
valid for small $r\kappa$, where $V$ is the particle volume. By equating this to the cross-section under the Born approximation $\sigmatotal=(nV)^2\sigma$, where $n$ is the number density of nucleons, we obtain
\begin{equation}
  \label{eq:13}
  \kappa^2 = \pm n\sqrt{4\pi\sigma}.
\end{equation}
Equivalently, and as in the field of neutron optics, we may parameterise by a `critical wavelength' $\lambda_c=2\pi/\kappa=\sqrt{\pi/n a_s}$ where $\sigma=4\pi a_s^2$.

\subsection{Decoherence in a matter-wave interferometer}
A full phase-space treatment is detailed elsewhere~\cite{Bateman:2013near}. We use the result for decoherence induced by isotropic elastic scattering including recoil; while this was originally derived for the case of Rayleigh scattering of black-body radiation, it is parameterised only in terms of a momentum and is therefore valid for the isotropic elastic scattering considered here:
\begin{equation}
  \label{eq:15}
  f(x)=1-\int_0^\infty\frac{\gamma(k)}{\Gamma}\left[\frac{\text{Si}(2kx)}{kx}-\text{sinc}^2(kx)\right] dk,
\end{equation}
where $\text{sinc}(x) \equiv \sin(x)/x$ and $\text{Si}(x) \equiv \int_0^x\text{sinc}(x')dx'$ is the Sine integral. It is possible to treat the full spectrum $\gamma(k)$ of incident $\chi$ particles; however, for our purposes it is sufficient to take the spectrum of incident wavenumbers to be narrow: $\gamma(k)=\Gamma\, \delta(k-k_0)$, where $k_0=m_\chi \bar{v}_0/\hbar$ is the typical wavenumber and $\bar{v}_0 \equiv v_0 c\sim 10^{-3} c$. While the exact value of the argument to this resolution function $f(x)$ in the expression for decoherence depends on the time scales and on the geometry of the interferometer experiment, it is very close to the grating spacing $\Lambda$.

The number of expected events is the flux of $\chi$ particles multiplied by the effective cross-section and the interaction time:
\begin{equation}
  \label{eq:16}
  W=\frac{\rho_\chi}{m_\chi}\bar{v}_0\,\sigma_\text{eff}\,\tau
  =\frac{3.6}{2\pi\hbar}\, \frac{\rho_\chi}{m_\chi}\bar{v}_0\,\sigma\,\Lambda^2 N^3\,\text{amu},
\end{equation}
where we have used $\sigma_\text{eff}=\sigma N^2$ which is valid for small particles. The characteristic time-scale for interference is given by the `Talbot time' $\tau_T=M\Lambda^2/(2\pi\hbar)$, where $M$ is the mass of the target particle; for the proposed interference experiment, $\tau=3.6\,\tau_T$.

\subsection{Penetration of Earth's atmosphere}
The refractive index model  implies that, for a sharp boundary between vacuum and dense matter, $\chi$ DM particles will be strongly reflected.  The critical wavelength for air at standard temperature and pressure ($n\approx 7.3\cdot 10^{26}\,\text{m}^{-3}$) is $\lambda_c\approx 4.6\,\upmu\text{m}$, which falls within the expected range for the $\chi$ particle's de Broglie wavelength $\lambda$: the expected refractive index $\eta$ could be close to unity or as large as 20 (20$i$) for an underlying attractive (repulsive) interaction.

In the attractive case, the very slow increase in atmospheric mass density on approach to the Earth's surface would strongly suppress Fresnel reflections; for the repulsive case, $\chi$ particles would penetrate only to a finite depth. Density fluctuations in the atmosphere, and the relatively large associated refractive index contrast, could lead to multiple reflections and hence to an effective finite penetration depth even in the attractive case. Such uncorrelated multiple events would reduce the kinetic energy of the incident $\chi$ particle.

In the case that $\chi$ particles do reach the Earth's surface, it may be possible (although challenging) to create a wavelength-scale structure of varying material densities which uses multiple reflections to engineer penetration of the particles into a dense solid.  With a such a device, $\chi$ particles could be focused and injected into an experimental vacuum chamber in which detection could be performed much as in the space-based experiment proposed in the main text.

\bibliographystyle{./apsrev}
\bibliography{./bibliography-arXiv}

\end{document}